\documentclass[printer]{aa}

\usepackage{amsmath}
\usepackage[varg]{txfonts}
\usepackage{float}

\begin{document}

\title{The formation and evolution of reconnection-driven, slow-mode shocks in a partially ionised plasma}

\author{A. Hillier\inst{1}, S. Takasao\inst{2} and N. Nakamura\inst{2}}

\institute{Department of Applied Mathematics and Theoretical Physics, University of Cambridge, Wiberforce Road, Cambridge CB3 0WA, UK\\ \email{ah826@cam.ac.uk}
\and
Kwasan and Hida Observatories, Kyoto University}

\titlerunning{Slow-mode shocks in partially ionised plasma}
\authorrunning{Hillier et al.}

\abstract
{The role of slow-mode magnetohydrodynamic (MHD) shocks in magnetic reconnection is of great importance for energy conversion and transport, but in many astrophysical plasmas the plasma is not fully ionised.
In this paper, we use numerical simulations to investigate the role of collisional coupling between a proton-electron, charge-neutral fluid and a neutral hydrogen fluid for the one-dimensional (1D) Riemann problem initiated in a constant pressure and density background state by a discontinuity in the magnetic field.
This system, in the MHD limit, is characterised by two waves. The first is a fast-mode rarefaction wave that drives a flow towards a slow-mode MHD shock wave.
The system evolves through four stages: initiation, weak coupling, intermediate coupling, and a quasi-steady state.
The initial stages are characterised by an over-pressured neutral region that expands with characteristics of a blast wave.
In the later stages, the system tends towards a self-similar solution where the main drift velocity is concentrated in the thin region of the shock front.
Because of the nature of the system, the neutral fluid is overpressured by the shock when compared to a purely hydrodynamic shock, which results in the neutral fluid expanding to form the shock precursor.
Once it has formed, the thickness of the shock front {is} proportional to $\xi_{\rm i}^{-1.2}$, which is a smaller exponent than would be naively expected from simple scaling arguments.
One interesting result is that the shock front is a continuous transition of the physical variables of sub-sonic velocity upstream of the shock front (a c-shock) to a sharp jump in the physical variables followed by a relaxation to the downstream values for supersonic upstream velocity (a j-shock). 
The frictional heating that results from the velocity drift across the shock front can amount to $\sim 2$ per cent of the reference magnetic energy.
}

   \maketitle

\section{Introduction}\label{INTRO}

Magnetic reconnection is the change in the connectivity of a magnetic field and is responsible for violent energy release in many space and astrophysical systems.
There are two key physical processes that are believed to be important in the energy conversion between magnetic energy and fluid (including thermal and kinetic) energy in magnetic reconnection: Joule heating and the work of the magnetic field on the fluid post-reconnection.
The latter of these mechanisms was proposed by \citet{PET1964}, {and was shown to be highly effective at converting the magnetic energy to kinetic and internal energies.
This is the case because this model was able to produce fast magnetic reconnection (roughly speaking the ratio of the inflow to outflow velocity is approximately 0.01 -- 0.1 and only weakly dependent on the Lundquist number) as a result of the presence of standing slow-mode shocks created by magnetic} field relaxation.
\citet{PRFO1986} developed a general class of steady-state two-dimensional (2D) reconnection models that included both the diffusion-dominated reconnection region and the convective region that surrounds it.
They found that both the \citet{PET1964} and \citet{SONN1970} models were limiting cases of the general class of models they found.
One key unifying feature of all the models was the presence of slow shocks associated with the relaxation of the magnetic field in the reconnection outflow region.
Nevertheless, it should be noted that gradients and rotation in the magnetic field can produce rotational discontinuities \citep{PET1967}.

In this paper we are interested in the oblique form of the slow-mode magnetohydrodynamic (MHD) shock known as the switch-off shock; the magnetic field is more perpendicular to the shock front downstream of the shock than upstream.
This is associated with the removal of the component of the magnetic field parallel with the shock front, but the component perpendicular to the shock front does not change owing to the requirement for $\nabla \cdot
\mathbf{B}$.
This results in a decrease in magnetic energy across the shock front that is balanced by an increase in both the thermal and kinetic energies of the fluid.
The abrupt change in direction of the magnetic field results in the acceleration of the fluid with a strong component of the velocity parallel to the shock front.
This can be viewed as analogous to the reconnection jet in the reconnection models cited in the previous paragraph.

Though there is still great debate about the role of slow-mode shocks in creating fast magnetic reconnection, their importance in driving observed dynamics of the solar atmosphere should not be underestimated.
Solar flares, which are huge energy releases in the solar atmosphere, are one of the observational signatures of magnetic reconnection in the solar corona \citep[for a recent review see, for example,][]{SHIB2011}. 
In the case of flare reconnection, thermal conduction  also plays an important role for energy transport and results in isothermal slow-mode shocks forming \citep[e.g.][]{LONG2010, TAKA2015}.

Another example that highlights the importance of reconnection-driven, slow-mode shocks was provided by \citet{TAKA2013}, who performed a detailed 2D numerical investigation of reconnection associated with emerging magnetic flux in the solar atmosphere. The main target of the work was to understand the role of magnetic reconnection in jet formation as a result of reconnection low down in the solar atmosphere where dynamically the energy release can have only limited consequences.
The results from this study showed that the slow-mode MHD shock is crucial for the transport of energy released by the reconnection to areas of the solar atmosphere where they can be of greater dynamic importance.
This example also highlights the great importance shocks can play in the transport of energy from reconnection to areas in the solar atmosphere 

The connection between shock physics, magnetic fields, and partially ionised plasma is of great importance in many astrophysical systems.
One key area of this study has been the study of shocks in the interstellar medium (ISM).
Fast-mode MHD shocks, where both the gas and magnetic field are compressed by the shock, in this weakly ionised medium involve a complex coupling between the magnetic field and the neutral gas.
Low ionisation of the medium means that the shock front can become a continuous transition over a large region \citep[known as a c-shock; ][]{DRAINE1980}; this allows radiative cooling to more efficiently cool the gas as it shocks, significantly reducing the temperature of the post-shock region and, as a result, significantly reducing the disassociation of molecules resulting from the shock \citep{CHER1987}.
However, certain solutions based on sufficiently large upstream velocities were found that, in spite of the continuous transition of the ionised fluid, the neutral gas would shock (known as a j-shock).
These solutions were normally associated with weak magnetic field strengths as this meant the magnetic field could not work to smooth out the shock front \citep{DRAINE1980}.
\citet{CHER1982} was able to use steady-state shock solutions to model molecular line intensities and predict the magnetic field strength of BN-KL.
For a review on this subject, see, for example, \citet{DRAINE1993}.

Other studies focussing on how aspects of magnetic reconnection other than its associated shocks are influenced by partial ionisation of the host plasma have lead to some very interesting results.
The current sheet structure  for a Harris-type current sheet under the influence of partial ionisation, through the one-fluid, ambipolar diffusion approximation, was found to form a power law with the magnetic field scaling as $B\propto x^{1/3}$ \citep{BRAN1994}.
\citet{ARB2009} extended this study to include a guide field in the current sheet showing that the current sheet develops into a $\mathbf{J} \times \mathbf{B}=0$ with a thickness ($l_{\rm new}$) that can be estimated as $l_{\rm new}=l_{\rm old}B_{\rm g}/B_{\rm ex}$, where $B_{\rm g}$ and $B_{\rm ex}$ are the original guide and external magnetic fields, respectively \citep{SINGH2015}.
A similar problem related to the current sheet thickness of the Kippenhahn-Schl\"{u}ter prominence model under ion-neutral drift was investigated by \citet{HILL2010}, finding that the thickness of the current sheet under ambipolar diffusion would become $l_{\rm new}=l_{\rm old}B_{\rm hor}/B_{\rm ver}$, where $B_{\rm hor}$ and $B_{\rm ver}$ are the original horizontal and vertical magnetic fields.

The reconnection process has also been shown to be modified by the influence of partially ionised and weakly ionised plasma.
\citet{SAKAI2008} and \citet{SAKAI2009} studied how reconnection in the penumbra of sunspots may be influenced by the low ionisation of the plasma finding that weak flows of the neutral hydrogen in the photosphere could greatly enhance the rate of reconnection in penumbral filaments.
\citet{SINGH2015} extended the fractal reconnection model of \citet{SHIB2001} to include the physics of collisional coupling of a partially ionised plasma in the strongly coupled, intermediately coupled, and weakly coupled regimes through application of the growth rates for the tearing instability derived by \citet{ZWEI1989}.
On application to the solar atmosphere they found that the fractal reconnection process would have to cascade down through all these levels to reach the kinetic scales that offer the possibility of fast magnetic reconnection.
\citet{LEAKE2012}  and \citet{LEAKE2013} investigated, through multi-fluid simulations, the coupling and decoupling of ion and neutral fluids in a reconnecting current sheet, finding that the reconnection inflow became decoupled, but the high-velocity reconnection jet was coupled.
They also found that for Lundquist numbers smaller than for those required for the plasmoid instability \citep[e.g.][]{huang2012} that the reconnection rate would become independent of Lundquist number.
{Looking at the possibility for the plasmoid instability to lead to Hall-mediated reconnection in various astrophysical bodies, \citet{VEK2013} found that a prime site for the plasmoid instability leading to fast, bursty reconnection would be in the weakly ionised medium of protostellar disks. 
It has been shown that the development of the plasmoids in a reconnection region leads to the dynamic formation of a multitude of slow shocks associated with the formation, movement, and merger of the plasmoids \citep{TANU2001, MIE2012, SHIB2015}. This finding highlights how the formation of plasmoids in a partially ionised plasma is intrinsically linked to the nature of shocks in that medium.}

As yet, the nature of slow-mode reconnection shocks in a partially ionised plasma are yet to be understood.
In this paper we look at a simple one-dimensional (1D) model that captures the necessary physics of the reconnection shock system but solves for the evolution of a neutral and ionised fluid that are coupled by collisions between the species.

\section{Solving the equations for a partially ionised plasma}

We investigate the dynamics of a neutral fluid and a charge-neutral, fully collisionally coupled ion electron plasma.
The equations governing the neutral fluid are written as
\begin{align}
\frac{\partial\rho_{\rm n}}{\partial t}+\nabla\cdot(\rho_{\rm n}\mathbf{v}_{\rm n})=&0 \\
\frac{\partial}{\partial t}(\rho_{\rm n}\mathbf{v}_{\rm n})+\nabla\cdot(\rho_{\rm n}\mathbf{v}_{\rm n}\mathbf{v}_{\rm n}+P_{\rm n}\mathbf{I})=& -\alpha_{\rm c}(T_{\rm n},T_{\rm p})\rho_{\rm n}\rho_{\rm p}(\mathbf{v}_{\rm n}-\mathbf{v}_{\rm p})\label{n_mom}\\
\frac{\partial e_{\rm n}}{\partial t}+\nabla\cdot[\mathbf{v}_{\rm n}(e_{\rm n}+P_{\rm n})]=&  \label{n_en}\\
-\alpha_{\rm c}(T_{\rm n},T_{\rm p})\rho_{\rm n}\rho_{\rm p}&\left[\frac{1}{2}(\mathbf{v}_{\rm n}^2-\mathbf{v}_{\rm p}^2)+3R_{\rm g}(T_{\rm n}-T_{\rm p})\right]\nonumber\\
e_{\rm n}=&\frac{P_{\rm n}}{\gamma-1}+\frac{1}{2}\rho_{\rm n} v_{\rm n}^2.
\end{align}
The equations governing the ionised fluid are written as
\begin{align}\label{MHD}
\frac{\partial\rho_{\rm p}}{\partial t}+\nabla\cdot(\rho_{\rm p}\mathbf{v}_{\rm p})=& 0 \\
\frac{\partial}{\partial t}(\rho_{\rm p}\mathbf{v}_{\rm p})+\nabla\cdot\left(\rho_{\rm p}\mathbf{v}_{\rm p}\mathbf{v}_{\rm p}+P_{\rm p}\mathbf{I}-\frac{\mathbf{BB}}{4\pi}\right.&+\left.\frac{\mathbf{B}^2}{8\pi}\mathbf{I}\right)=\label{p_mom} \\
\alpha_{\rm c}& (T_{\rm n},T_{\rm p})\rho_{\rm n}\rho_{\rm p}(\mathbf{v}_{\rm n}-\mathbf{v}_{\rm p})\nonumber
\end{align}
\begin{align}
\frac{\partial}{\partial t}\left( e_{\rm p}+\frac{B^2}{8\pi} \right)+ \nabla\cdot & \left[\mathbf{v}_{\rm p}(e_{\rm p}+P_{\rm p})+\frac{c}{4\pi}\mathbf{E}\times\mathbf{B}\right]=  \label{p_en}\\
 \alpha_{\rm c}&(T_{\rm n},T_{\rm p})\rho_{\rm n}\rho_{\rm p}\left[\frac{1}{2}(\mathbf{v}_{\rm n}^2-\mathbf{v}_{\rm p}^2)+3R_{\rm g}(T_{\rm n}-T_{\rm p})\right]\nonumber\\
\frac{\partial \mathbf{B}}{\partial t}+\nabla \times (\mathbf{v}_{\rm p}\times \mathbf{B})= & 0\\
e_{\rm p}=&\frac{P_{\rm p}}{\gamma-1}+\frac{1}{2}\rho_{\rm p} v_{\rm p}^2\\
\nabla\cdot\mathbf{B}=& 0.\label{DB}
\end{align}
We take both fluids to be ideal gases following the relations $P_{\rm n}=\rho_{\rm n}R_{\rm g}T_{\rm n}$ and $P_{\rm p}=2\rho_{\rm p}R_{\rm g}T_{\rm p}$, respectively.
In this formulation $\alpha_{\rm c}\rho_{\rm \alpha}=\nu_{\rm \alpha\beta}$, which is the collision frequency of species $\alpha$ on species $\beta$ where the subscripts $\rm \alpha$ and $\rm \beta$ denote either $\rm n$ or $\rm p$. 
We formulated these {equations} with {extensive} reference to previous numerical codes that were used to investigate partially ionised plasma (see Appendix \ref{Appen1} for more details.)

The key difference between this and single fluid models, i.e. those that approximate the ion-neutral coupling through a modified Ohm's law \citep[e.g. ][]{BRAG1965}, is that we solve each fluid separately, and couple them through the collisional coupling terms.
These terms can be found as the source terms on the RHS of Equations \ref{n_mom}, \ref{n_en}, \ref{p_mom}, and \ref{p_en}.
The code used to study this problem is the (P\underline{I}P) code, which we developed to study the influence of partial ionisation on the dynamics of magnetised fluids.
A basic description of the code, the full equations it can solve and tests of the collisional coupling terms are presented in Appendix \ref{Appen1}.
For this study, as we are interested in a shock problem, we use an HLLD scheme \citep{MIYO2005} because this was found to be stable down to very low ionisation fractions and plasma $\beta$ values.

\section{The model under consideration}

In this study, we extend the model of slow-shocks formed as a result of magnetic reconnection proposed by \citet{PET1964}. {We examine the changes that occur when} we look at plasmas of differing neutral fraction.
The equations are normalised such that the density $\rho_{\rm tot}=1$, the collision frequency as determined by the bulk fluid density is $\nu=\alpha_{\rm c}(T_0)\rho_{\rm tot}=1=\tau^{-1}$ and the bulk Alfv\'{e}n velocity $V_{\rm A}=B_0/\sqrt{\rho_{\rm tot}}=1;$  the normalisation of the magnetic field is used such that $B_0=B_{\rm norm}/\sqrt{4\pi}$.
The value of $\alpha_{\rm c}(T_0)$ is normalised to one, but $\alpha_{\rm c}(T_{\rm n},T_{\rm p})$ depends on the temperature as follows:
\begin{equation}
\alpha_{\rm c}(T_{\rm n},T_{\rm p})=\alpha_{\rm c}(T_0)\sqrt{\frac{T_{\rm n}+T_{\rm p}}{2T_0}}
.\end{equation}
Physically this means that a bulk Alfv\'{e}n wave approximately becomes coupled to both fluids after travelling through a unit length as defined by $L_{\rm norm}=V_{\rm A}/\nu$.
This normalisation means that the plasma $\beta$ is defined as $\beta=P_{\rm tot}/(B_0^2/2)$.

The initial conditions in normalised units are as follows:
\begin{align}
B_x=&0.3B_0\\
B_y=&\begin{cases} -B_0 & \mbox{if $x > 0$};\\ B_0 & \mbox{if $x < 0$}\end{cases}\\
\rho_{\rm n}=&\xi_{\rm n} \rho_{\rm tot}\\
\rho_{\rm p}=&\xi_{\rm i} \rho_{\rm tot}=(\xi_{\rm n}-1) \rho_{\rm tot}
\end{align}
\begin{align}
P_{\rm n}=&\frac{\xi_{\rm n}}{\xi_{\rm n}+2\xi_{\rm i}}P_{tot}=\frac{\xi_{\rm n}}{\xi_{\rm n}+2\xi_{\rm i}}\beta \frac{B_0^2}{2}\\
P_{\rm n}=&\frac{2\xi_{\rm i}}{\xi_{\rm n}+2\xi_{\rm i}}P_{tot}=\frac{2\xi_{\rm i}}{\xi_{\rm n}+2\xi_{\rm i}}\beta \frac{B_0^2}{2},
\end{align}
where $\xi_{\rm n}$ and $\xi_{\rm i}$ are the neutral and ion fractions, respectively.
We are assuming that the two fluids are in thermal equilibrium at the start of the calculation.
It should be noted that this angle of magnetic field {and plasma $\beta$ values used are} known to produce a slow-mode, switch-off shock and not a purely hydrodynamic shock \citep{TAKA2015}.
The key parameters that we investigate are the ionisation fraction and plasma beta (ratio of gas to magnetic pressure) of the system, and so that these parameters are varied between simulations, the other parameters are kept the same through all the simulations.
We take the adiabatic index of $\gamma=5/3$.
The 1D model under consideration here has been shown to give a good approximation of the dynamics of more complex reconnection simulations 
\citep{TAKA2013}.

As a result of the symmetry around $x=0$ of the situation we are studying, we can take only {one-half} the computational domain and set the boundary at $x=0$ as a reflective boundary formulated {in such a way that} the magnetic field can penetrate {the} boundary.
The boundary at $x=x_{\rm max}$ is an open boundary.
As this is a 1.5D simulation the derivatives in the $y$ direction are taken to be $0$ and we do not include any velocity or magnetic field component in the $z$ direction.
The spatial resolution in the $x$ direction of the simulations is $\Delta x=0.1 \xi_{\rm i0}/\xi_{\rm i}$, where $\xi_{\rm i0}=0.1$, unless otherwise stated.

\subsection{The reference MHD solution}\label{ref_MHD}

Though the MHD solution of this system has been well investigated, to provide sufficient context for the results we present a reference MHD solution.
This is obtained by solving the system described above using Equations \ref{MHD} to \ref{DB}, where the source terms on the RHS of these equations are neglected (i.e. we use the ideal MHD equations{ and we examine only the ions}).

Before we introduce the results, there is one key point that should be stressed.
The equations used here are the ideal MHD equations, and these equations have the special property of having no inherent scale of reference.
This is different from the case where gravity is involved, which naturally has the pressure-scale height as its intrinsic length scale, or, as is relevant for this study, when partially ionised plasmas are considered as they have timescales associated with their collisional coupling.
Therefore, when looking at an ideal MHD system, it is no surprise that it displays self-similarity, and as such we can present only one snapshot of the system here and know that it is representative of the system at all times. This does not apply to all MHD systems, but is a common feature of expanding shock systems \citep[e.g. the hydrodynamic Sedov-Taylor point explosion][]{SED1959}.
\begin{figure*}[ht]
\centering
\includegraphics[width=13.5cm]{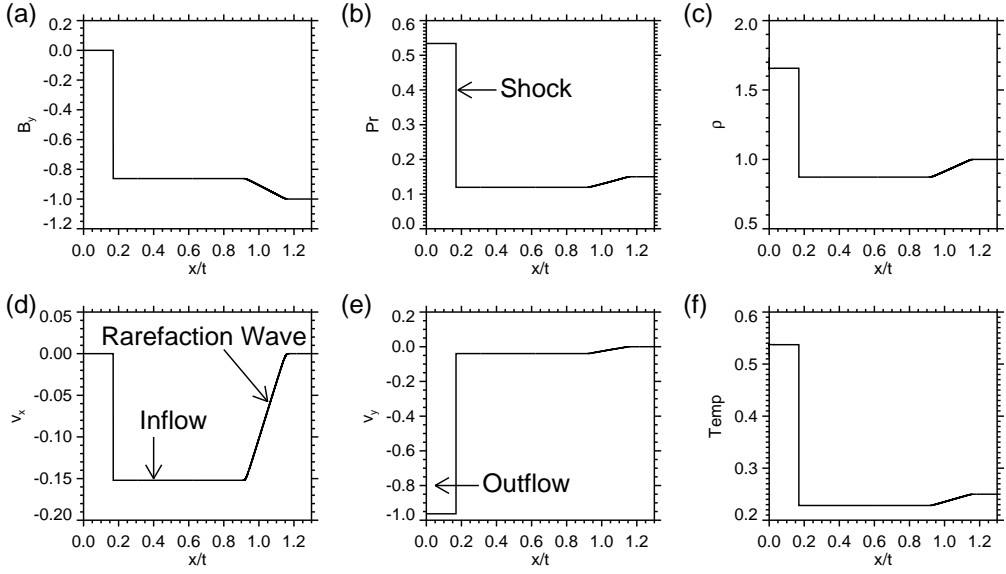}
\caption{Spatial distribution in the $x$ direction of $B_{\rm y}$ (a), gas pressure (b),  density (c), $v_{\rm x}$ (d), $v_{\rm y}$ (e), and  temperature (f) for the ideal MHD simulation. There is a fast-mode rarefaction wave at approximately $x/t=1$ that drives a flow of material towards the shock front at $x/t=0.175$}
\label{ref_MHD_shock}
\end{figure*}

Figure \ref{ref_MHD_shock} gives a snapshot of the distributions (going clockwise from the top left) of the $B_y$ magnetic field (the $B_x$ magnetic field is always constant to satisfy $\nabla \cdot \mathbf{B}=0$ and $B_{\rm x}=0.3$ in this case), gas pressure, density, temperature, $v_{\rm y}$ and $v_{\rm x}$.
The values at $x/t=1.3$ are those of the initial conditions of the simulation.
In this system, the relaxation of the magnetic field leads to the formation of two nonlinear waves: a fast-mode rarefaction wave and a slow-mode shock wave.
The fast-mode rarefaction wave is responsible for driving the inflow of material towards the shock front.
It is characterised by a linear profile in the transition between the pre- and post-wave values for $B_{\rm y}$, pressure, density, $v_{\rm x}$ and $v_{\rm y}$.

Post shock there is a high-velocity jet in the $y$ direction; this can be seen as analogous to a reconnection jet.
The post-shock velocity becomes $v_{\rm x}=0$ in the simulation frame, as does the post-shock $y$-component of the magnetic field ($B_{\rm y}$).
The shock jump conditions for this particular MHD shock (in the shock reference frame) are \citep[see, for example, ][]{GOED2004} written as
\begin{align}
B_{\rm xU}=&B_{\rm xD}=B_{\rm x}\\
\rho_{\rm U}v_{\rm xU}=&\rho_{\rm D}v_{\rm xD} \label{cont}\\
v_{\rm xU}B_{\rm yU}-v_{\rm yU}B_{\rm x}=&-v_{\rm yD}B_{\rm x}\label{ind}\\
\rho_{\rm U}v_{\rm xU}^2+p_{\rm U}+\frac{B_{\rm yU}^2}{2}=&\rho_{\rm D}v_{\rm xD}^2+p_{\rm D}\label{x_mom}\\
\rho_{\rm U}v_{\rm xU}v_{\rm yU}-B_{\rm x}B_{\rm yU}=&\rho_{\rm D}v_{\rm xD}v_{\rm yD}\label{y_mom}\\
\left(\frac{1}{2}\rho_{\rm U}(v_{\rm xU}^2+v_{\rm yU}^2)+\frac{\gamma}{\gamma-1}p_{\rm U}\right)&v_{\rm xU}+B_{\rm yU}(v_{\rm xU}B_{\rm yU}-v_{\rm yU}B_{\rm x})= \\
\frac{1}{2}\rho_{\rm D}(v_{\rm xD}^2&+v_{\rm yD}^2)v_{\rm xD}+\frac{\gamma}{\gamma-1}p_{\rm D}v_{\rm xD}\nonumber,
\end{align}
where the subscripts $\rm U$ and $\rm D$ denote the upstream and downstream quantities.
We also need to supplement this with a condition that the entropy must increase across the discontinuity, i.e.
\begin{equation}
\rho^{-\gamma}_{\rm U}P_{\rm U}\le \rho^{-\gamma}_{\rm D}P_{\rm D}
.\end{equation}
Equation (\ref{cont}) tells us that the ratio of the two densities is given by the ratio of $v_{\rm xU}/v_{\rm xD}$, which in this case is approximately $1.9$.
By combining Equations (\ref{cont}), (\ref{ind}), and (\ref{y_mom}) we can derive the relation $(v_{\rm yD}-v_{\rm yU})^2=\Delta v_{\rm y}^2=V_{\rm AU}^2$, where $V_{\rm AU}^2$ is the upstream Alfv\'{e}n velocity (this was checked to be true in the simulations to an accuracy of $0.0001$\,per cent).
As the upstream velocity in the $y$ direction is likely to be significantly smaller than the Alfv\'{e}n velocity, it can be said that approximately the post-shock jet is travelling at the Alfv\'{e}n velocity.
Then by combining Equations (\ref{cont}) and (\ref{x_mom}) we can show that $\Delta p =-\rho_{\rm U}V_{\rm xU}\Delta v_{\rm x} + B_{\rm y}^2/2$, which implies that the gas pressure has to increase to match both the dynamic and magnetic pressure drop, making it possible for pressure increases that are much larger than those possible for purely hydrodynamic shocks, as both of these terms can be significantly larger than $p_{\rm U}$ in this model for low plasma $\beta$.

\section{Temporal evolution}

In this section we take one specific set of parameters and study in detail the evolution of the formation of the shock and rarefaction waves through time, focussing on times that are smaller than the coupling time to those that are significantly longer.
For this part of the study, we take $\xi_{\rm n}=0.9$ and $\beta=0.3$.

\subsection{Initiation}

The initialisation of the dynamics is dominated by the evolution of the magnetised fluid; this is simply because of the initial conditions (i.e., only the magnetic field is out of equilibrium).
When the system is initialised, as with the ideal MHD simulation in Sect. \ref{ref_MHD}, a fast-mode rarefaction wave and a slow-mode shock form in the ionised fluid.
For these initial conditions, the bulk Alfv\'{e}n velocity of the fluid is normalised to 1, therefore the Alfven velocity with respect to the ionised fluid is larger, in this case $V_{\rm Ap}=\frac{V_{\rm A}}{\sqrt{\xi_{\rm i}}}\sim \sqrt{10}V_{\rm A}$.
This results in a fast-mode rarefaction wave travelling only in the ionised fluid away from the position of the original discontinuity in the magnetic field at approximately the ion Alfv\'{e}n speed.
This wave is accelerating the ionised fluid towards a slow-mode MHD shock.
Downstream of this shock, there is a jet of the ionised fluid that is travelling at a velocity of a few times the bulk Alfv\'{e}n speed.
However, as there is no force on the neutrals that can accelerate them in that direction until the coupling has taken effect, their velocity in the $y$ direction is $0$.
This large velocity of the ionised fluid jet in the $y$ direction, in conjunction with the initially $0$ velocity of the neutral fluid in this direction, results in a large velocity difference $v_{\rm D}$ in the jet region.

\begin{figure*}[ht]
\centering
\includegraphics[width=15.cm]{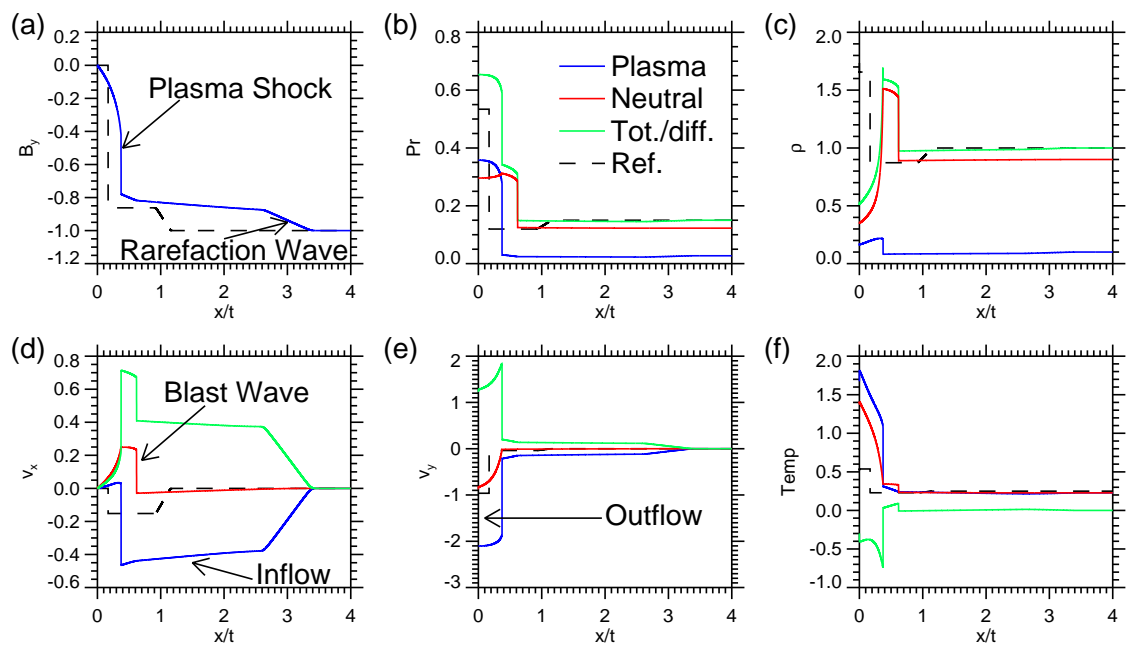}
\caption{Spatial distribution in the $x$ direction of $B_{\rm y}$ (a),  gas pressure (b), density (c), $v_{\rm x}$ (d), $v_{\rm y}$ (e), and temperature (f) for the neutral (red) and ionised (blue) fluids at time $t=1 \tau$. The green line indicates the total (pressure and density) or the difference ($x$ and $y$ velocities and temperature) for the two fluids. The dashed black line shows the reference MHD solution. The ionised plasma has taken on characteristics that are similar to the ideal MHD solution (but with faster wave speeds); the neutrals, however, are undergoing a violent expansion similar to that of a 1D point explosion.}
\label{initial_PIP_shock}
\end{figure*}

\begin{figure*}[ht]
\centering
\includegraphics[width=6.cm]{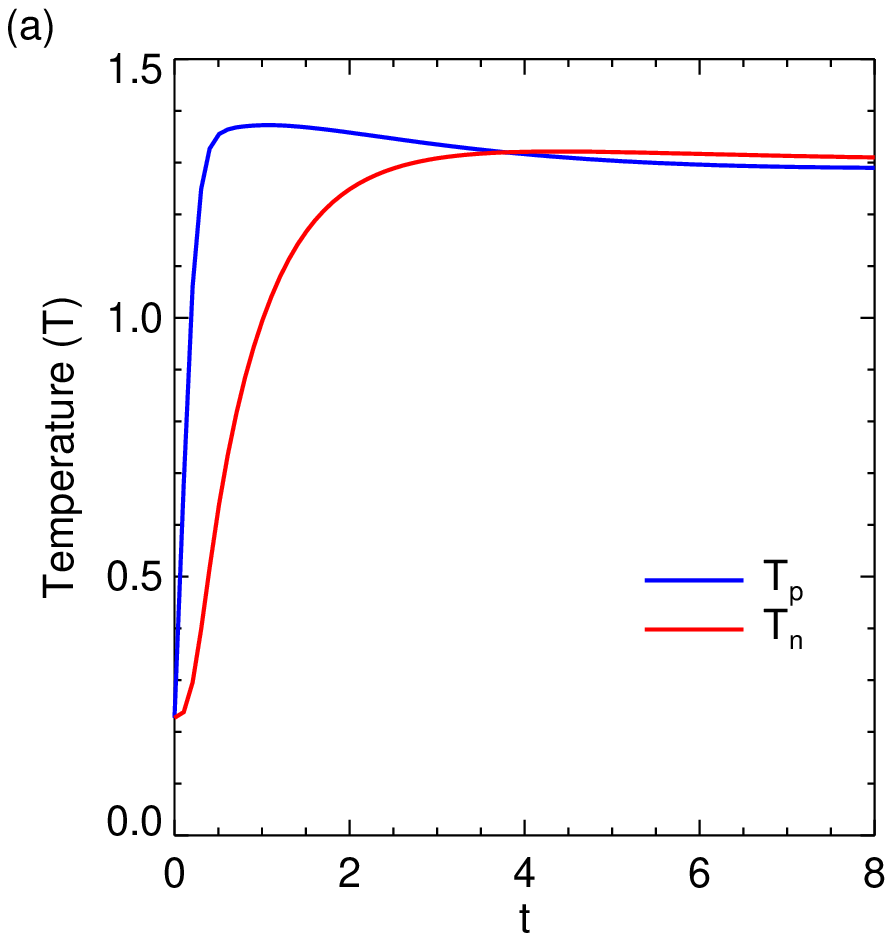}
\includegraphics[width=6.cm]{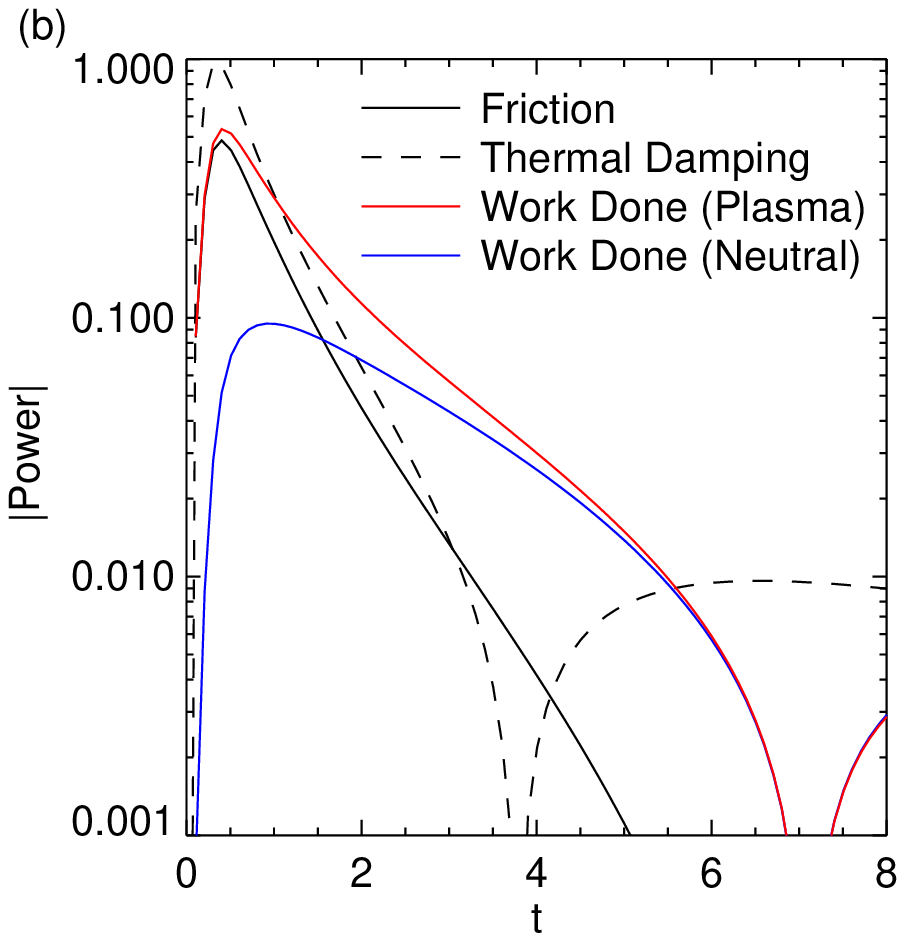}
\caption{Panel (a) indicates the temporal evolution of the temperature and panel (b) indicates the magnitude of the heating terms relating to collisional coupling at $x=0$ for a timescale that follows the transition from the initiation to the weak coupling regimes (see Sect. \ref{weak_coup}). The normalisation of the power is $B_0^2 \nu$.  Because the temperatures cross at $t \sim 3.5\tau$ the thermal damping term become 0 at this point then reverse sign (i.e. the ionised fluid is losing heat to the neutral fluid before this time, but gaining heat from the neutral fluid after this point).}
\label{initial jet temp}
\end{figure*}

The large $v_{\rm D}$ in the jet region results in the nonlinear terms of the coupling (those {of order} $v_{\rm D}^2$) becoming much larger than those {of order} $v_{\rm D}$.
Therefore, the coupling of the fluids {comes about via} the nonlinear coupling terms in the energy equations.
The left-hand panel of Fig. \ref{initial jet temp} gives the temperature as a function of time at $x=0$ for both the ionised and neutral fluids.
Initially the temperature of the ionised fluid increases drastically followed more slowly by the neutral fluid after about $3.5\tau$ {(where $\tau=\nu^{-1}$)} the two fluids have approximately the same temperature.
The right panel shows the size of the heating (cooling) terms associated with frictional heating given by $\alpha_{\rm c}(T_{\rm n},T_{\rm p})\rho_{\rm n}\rho_{\rm p}(\mathbf{v}_{\rm n}-\mathbf{v}_{\rm p})^2$ (solid line) and thermal damping term that drives thermal equilibrium given by $3\alpha_{\rm c}(T_{\rm n},T_{\rm p})\rho_{\rm n}\rho_{\rm p}(p_{\rm n}/\rho_{\rm n}-p_{\rm p}/2\rho_{\rm p})$ (dashed line)  and the two terms for the work performed on the fluid by the drift velocity given by $|\alpha_{\rm c}(T_{\rm n},T_{\rm p})\rho_{\rm n}\rho_{\rm p}\mathbf{v}_{\rm n} \cdot (\mathbf{v}_{\rm n}-\mathbf{v}_{\rm p})|$ (blue line) and $|\alpha_{\rm c}(T_{\rm n},T_{\rm p})\rho_{\rm n}\rho_{\rm p}\mathbf{v}_{\rm p} \cdot (\mathbf{v}_{\rm n}-\mathbf{v}_{\rm p})|$ (red line), respectively.
The thermal damping term is equal in magnitude but opposite in sign for each fluid.
This highlights how the frictional heating term, which represents a non-reversible increase in entropy, forms a significant proportion of the heating in the early stages.

The neutral fluid presents a very interesting response to its initial heating.
The localised heating of this fluid results in an increase in pressure around $x=0$, but it remains at its initial value elsewhere. 
This means we have a highly localised, high-temperature region, which, other than the external plasma not being cold, results in a Sedov-Taylor like explosive expansion of the inner high-temperature neutral layer.
This can be seen from {the neutral expansion, driven by a blast wave, shown in panel (d) of} Fig. \ref{initial_PIP_shock}.

\subsection{Weak coupling}\label{weak_coup}

\begin{figure*}[ht]
\centering
\includegraphics[width=15.cm]{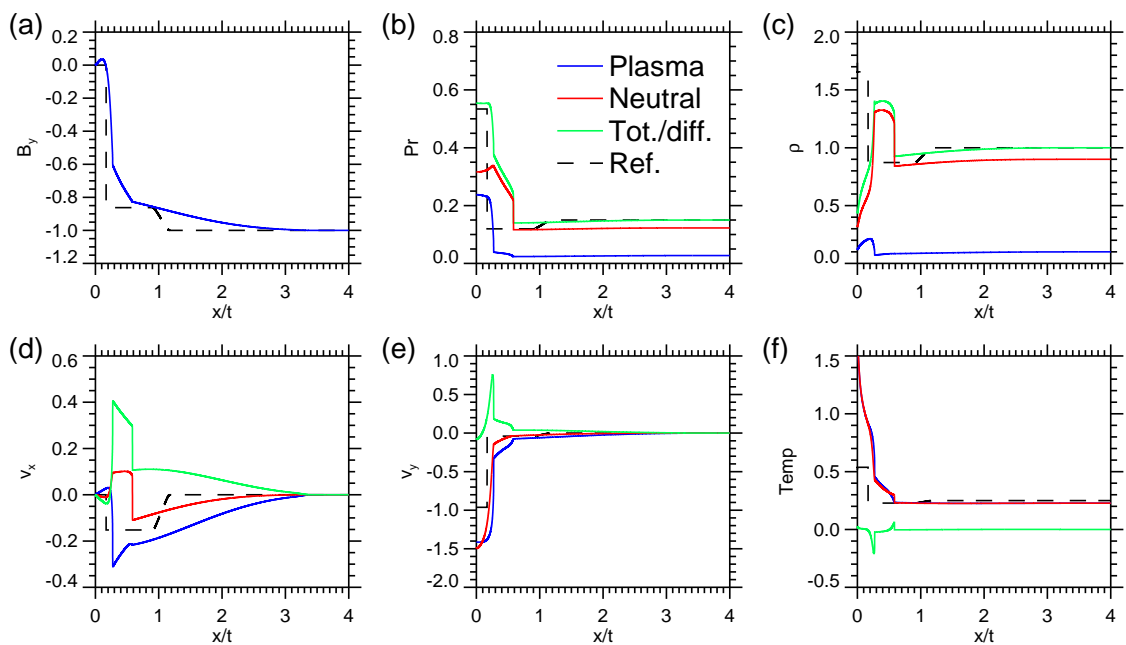}
\caption{Spatial distribution in the $x$ direction of $B_{\rm y}$ (a), gas pressure (b),  density (c), $v_{\rm x}$ (d), $v_{\rm y}$ (e), and temperature (f) for the neutral (red) and ionised (blue) fluids at time $t= 10\tau$. The green line indicates the total (pressure and density) or the difference ($x$ and $y$ velocities, and temperature) for the two fluids. The dashed black line indicates the reference MHD solution.}
\label{weak_PIP_shock}
\end{figure*}

Figure \ref{weak_PIP_shock} shows the system early in its evolution at a time ($10\tau$) that is an order of magnitude greater than that of the previous subsection.
As the coupling between the ions and neutrals is beginning to take effect, we can see that structures that are present in both fluids are beginning to form.
There are four components of the system: the rarefaction wave,  neutral shock, ion shock, and post-shock region.
All the regions in which flows are present display velocity drifts and these drifts are of approximately the same magnitude as the flows themselves.

By this time, both an inflow in the pre-shock region and post-shock jet are developing in the neutral fluid as a result of the coupling between the two fluids.
There is a shock in the neutral fluid that is propagating away from the origin in advance of the ionised plasma.
This is the continuation of the explosion in the neutral fluid described earlier.
However, as the {post neutral-shock region} is a prime {site} for frictional heating because of the large differences in velocity, it results in energy that is constantly added to this shock wave resulting in the peak in pressure that is some distance behind the shock front.
The expansion has resulted in a density depletion at the origin.

\subsection{Intermediate coupling}

\begin{figure*}[ht]
\centering
\includegraphics[width=15.cm]{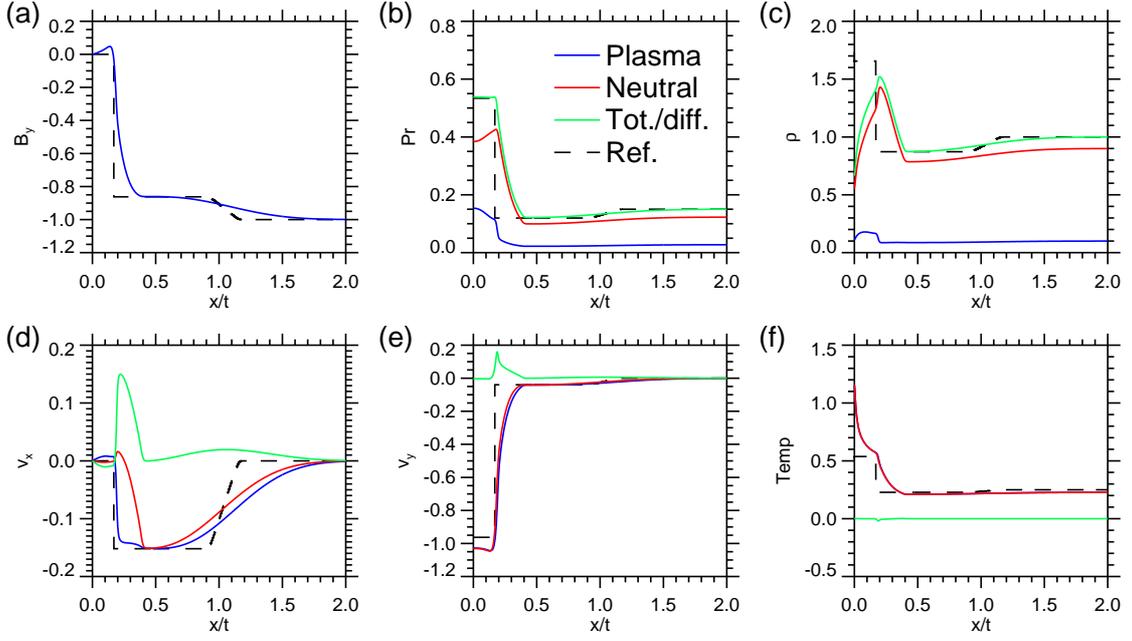}
\caption{Spatial distribution in the $x$ direction of $B_{\rm y}$ (a),  gas pressure (b), density (c), $v_{\rm x}$ (d), $v_{\rm y}$ (e), and temperature (f) for the neutral (red) and ionised (blue) fluids at time $t= 100\tau$. The green line indicates the total (pressure and density) or the difference ($x$ and $y$ velocities, and temperature) for the two fluids. The dashed black line indicates the reference MHD solution.}
\label{Intermediate_PIP_shock}
\end{figure*}

Figure \ref{Intermediate_PIP_shock} looks at the system at a time ($100\tau$) that is approximately an order of magnitude later than that of the previous subsection, and as such the dynamics of the two fluids show\ greater coupling.
The high pressure neutral region has sufficiently expanded so that the pressure at the front of this expansion has dropped to match that of the ambient pressure, and as such the shock in the neutrals has disappeared and is replaced with a smooth transition between the inflow region of neutral and ionised fluid and the jet region. 
It is in this smooth transition region, which represents a c-shock, that the high levels of drift velocity are found ($\sim 0.15V_{\rm A}$).
The two waves that we expect to form in this system, the slow-mode shock and fast-mode rarefaction wave, are visible, but are yet to form as completely distinct elements.

It can be seen that in the region of the c-shock, the neutral fluid is undergoing the velocity transition before the ionised fluid.
It is worth noting that this is a clear difference from the c-shocks studied that are relevant to the ISM.
This leads to the pre-heating in the shock front relative to the reference MHD solution.

\subsection{Quasi-self-similar state}\label{QSSS}

\begin{figure*}[ht]
\centering
\includegraphics[width=15.cm]{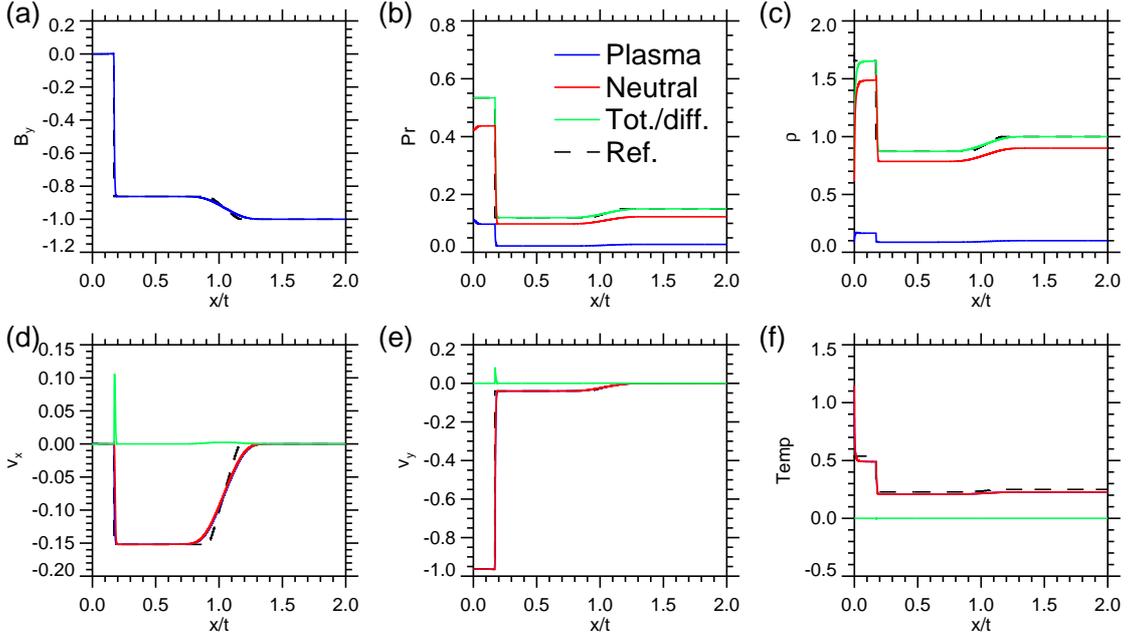}
\caption{Spatial distribution in the $x$ direction of $B_{\rm y}$ (a), gas pressure (b), density (c), $v_{\rm x}$ (d), $v_{\rm y}$ (e), and temperature (f) for the neutral (red) and ionised (blue) fluids at time $t=2000 \tau$. The green line indicates the total (pressure and density) or the difference ($x$ and $y$ velocities, and temperature) for the two fluids. The dashed black line indicates the reference MHD solution. The solution for the partially ionised case has reverted to that of the ideal MHD case outside of the wave fronts.}
\label{self_sim_PIP_shock}
\end{figure*}

Once the system has sufficiently evolved then it reaches a state that is approaching that of self-similarity.
Figure \ref{self_sim_PIP_shock} gives an example of this later stage at a time ($2000\tau$) that is a factor of 20 greater than that used in the previous subsection.
There are now two distinct waves in the system, the fast-mode rarefaction wave and the slow-mode shock.
The pre- and post-shock values of the physical quantities are basically constant and large values of the drift velocity are only found in the shock front.
As the drift velocity in the $x$ direction is positive, we can tell that the neutral fluid undergoes the shock transition before the ionised fluid, but for the post-shock jet the ionised fluid is accelerated first and then the neutral fluid is dragged with it.
The nonlinear coupling terms related to frictional heating are most important in this region.

It is worth considering what conditions are necessary to produce this steady-state shock.
Eventually, the rarefaction wave and  shock front decouple, and after this time the density, momentum, energy, and magnetic flux that leaves the rarefaction wave and as such composes the upstream conditions for the shock becomes almost constant in time (at least rapidly tends to such a state).
This results in the shock jump conditions that need to be satisfied at each timestep becoming the same, which results, when travelling in the shock reference frame, in a steady-state shock.
In the following, we provide an analysis of why this quasi-self-similar solution develops and what determines the end state of this system.

\begin{figure}[ht]
\centering
\includegraphics[width=6.5cm]{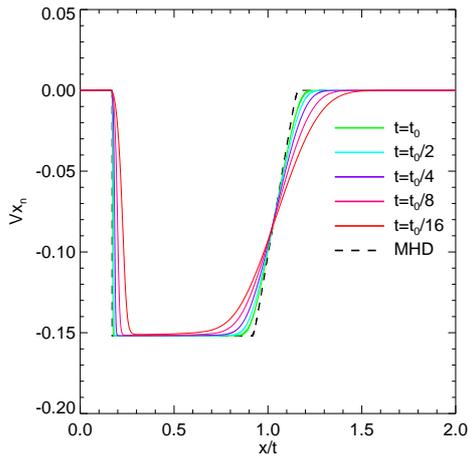}
\caption{Quasi-self-similar evolution of the shock-rarefaction wave system. To highlight how the system is tending towards self-similarity, the $x$-axis is normalised by the time $t$ of the snapshot of the simulation, therefore shows the velocity at which the waves are propagating  in normalised units. As the time is increased, the system tends towards a self-similar state. The reference time is taken at $t_0=156000\tau$ and the distributions for $t_0$, $t_0/2$, $t_0/4$, $t_0/8,$ and $t_0/16$ are plotted. The reference ideal MHD solution is indicated by the black dashed line.}
\label{QSS}
\end{figure}

Figure \ref{QSS} shows the results from a simulation where $\beta=0.3$ and $\xi_{\rm n}=0.99$.
This calculation has been run for {a} longer time (using $\Delta x =10$) specifically to analyse the evolution towards self-similarity.
The $x$-axis of this figure is the length scale in the $x$ direction divided by the time that the snapshot of the simulation was taken.
In this way, a feature travelling at a constant velocity always appears at the same point in the figure at all times.
Here we can see that the position of centroid of rarefaction wave and the start of the post-shock region evolve in a self-similar fashion (i.e. these points travel at a constant velocity).
The relative thickness of these two layers, however, does change; a detailed investigation of this is presented below.

\subsubsection{The effective thinning of the rarefaction wave}\label{rarefaction}

Figure \ref{QSS} shows the evolution of the rarefaction wave as the system heads towards a final state.
 In the velocity reference frame used, the thickness of the rarefaction wave decreases with time, tending towards the reference ideal MHD solution.
This effective thinning of the wave front is a result of the diffusive processes becoming less important with time when compared with the dynamic processes of the system as it evolves.
This can be simply modelled by taking the thickness of the rarefaction wave ($W_{\rm R}$) to be given by the sum of component that comes from the expansion of the system (${\rm A} t$) and diffusion of the system (${\rm B} t^{1/2}$), i.e.
\begin{equation}
W_{\rm R}={\rm A} t + {\rm B} t^{1/2}
,\end{equation}
where $t>0$.
Therefore,
\begin{equation}
\frac{W_{\rm R}}{t}={\rm A} + {\rm B} t^{-1/2}
,\end{equation}
which implies that a $t$ gets larger, the influence of the diffusive component becomes smaller meaning that
\begin{equation}
\frac{W_{\rm R}}{t} \xrightarrow{t\to\infty}{\rm A}
,\end{equation}
i.e. the system is heading towards a self-similar state at the rate of $t^{-1/2}$.
As a note, the physical meaning of $A$ is the velocity difference between the entry and exit points of the rarefaction wave for a fluid parcel that is given by the ideal MHD solution.
For this particular simulation, by calculating the Dopplershifted fast-mode wave speeds (in the simulation frame) upstream and downstream of the wave gives $V_{\rm fU}=1.150V_{\rm A}$ and $V_{\rm fD}=0.926V_{\rm A}$, therefore ${\rm A}=0.223V_{\rm A}$.


Another way of thinking about this relates to the drift velocities and relative frequencies of the momentum coupling as the rarefaction wave expands.
If we think about a packet of the neutral fluid and its acceleration by the rarefaction wave towards the shock front entering the wave at time $t_0$ and exiting at time $t_1$.
The amount of momentum transferred to the neutral fluid by collisions with the ionised fluid $(\rho_{\rm n}v_{\rm xn})_{\rm C}$ must be constant in time to give the same velocity, i.e.\begin{equation}
(\rho_{\rm n}v_{\rm xn})_{\rm C}=\int^{t_1}_{t_0}\alpha_{\rm c}(T_{\rm n},T_{\rm p})\rho_{\rm n}\rho_{\rm p}(v_{\rm xn}-v_{\rm xp})\partial t={\rm Const}.
\end{equation}
Therefore, as the time taken for the packet of neutral fluid ($t=t_1-t_0$) to cross the wave front gets larger, then the average velocity drift that the fluid packet experiences ($\langle{v_{\rm xn}-v_{\rm xp}}\rangle$) tends to zero.
As this happens, the nonlinear term involved with the dissipation of the kinetic energy associated with the drift velocity through frictional heating must also change as follows:
\begin{equation}
\int^{t_1}_{t_0}\alpha_{\rm c}(T_{\rm n},T_{\rm p})\rho_{\rm n}\rho_{\rm p}(v_{\rm xn}-v_{\rm xp})^2\partial t \xrightarrow{t\to\infty} 0
.\end{equation}
With this approach, we can see that two things are happening: 1) the system is tending towards a state of perfect coupling where the ratio of the dynamic timescale ($\tau_{\rm D}$) to the collisional coupling timescale $\tau_{\rm D} \nu \rightarrow \infty$ and 2) that the frictional heating term becomes $0$, which means that the nonlinear terms cannot change the state of the upstream shock conditions.

\subsubsection{Steady-state shock layer}

\begin{figure}[ht]
\centering
\includegraphics[width=6.5cm]{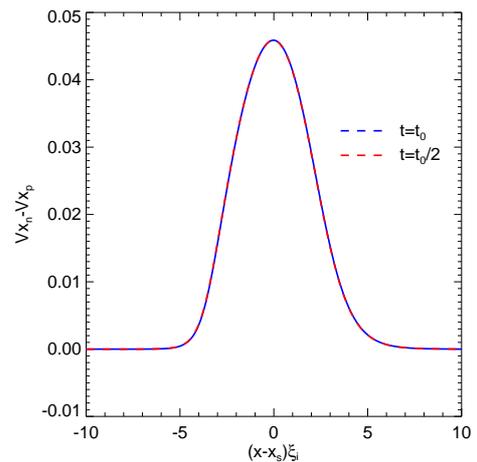} 
\caption{Drift velocity through the shock front for the simulation with $\beta=0.3$ and $\xi_{\rm i}=0.01$. The $x$-axis is shifted by $x_{\rm s}$, where $x_{\rm s}$ is taken as the point in the shock where the drift velocity is at its peak, so that both distributions are aligned. It is clear that the distribution of the drift velocities is the same at both times indicating that the shock front has reached a steady state. The reference time is taken at $t_0=120000\tau$.}
\label{steady_shock}
\end{figure}

Figure \ref{steady_shock} shows the drift velocity distribution across the shock front for the simulation where $\beta=0.3$ and $\xi_{\rm i}=0.01$.
The central position of the shock has been normalised as the position of the peak drift velocity.
The distribution is taken at two different times $t=60000\tau$ and $t=120000\tau$, but the distribution of the velocity drift across the shock is the same at these two times.
After a sufficient period of time the shock reaches a steady state.

It is worthwhile here to discuss the difference between the shock structure found for these calculations and those investigated for the ISM.
Because this study deals with a slow-mode shock and not a fast-mode shock as well as plasma $\beta$ values that are less than $1$, it is very natural that there are differences in the two approaches.
In the fast-mode shock case, the extra magnetic pressure that comes from the compression of the magnetic field results in the expansion of the ionised fluid layer, which couples to the neutral fluid and expands that layer.
In this case, as the slow-mode shock has a decrease in magnetic pressure that is partly balanced by an increase in gas pressure, it is the neutral fluid that is over-pressured by the shock and this region pushes forwards.


\section{Parameter dependence of shock structure}\label{depend}

In the previous section, we mainly focussed on the temporal evolution of one particular example.
Now we focus on the structure of the shock region once the system has reached the quasi-self-similar state as described in Section \ref{QSSS} and investigate the influence {of some} of the parameters of the system on the structure of the shock.
Keeping all other parameters the same, we investigate the two-dimensional parameter space of ion fraction $\xi_{\rm i}$ and plasma $\beta$.

\subsection{Ionisation fraction dependence}

\begin{figure*}[ht]
\centering
\includegraphics[width=6.5cm]{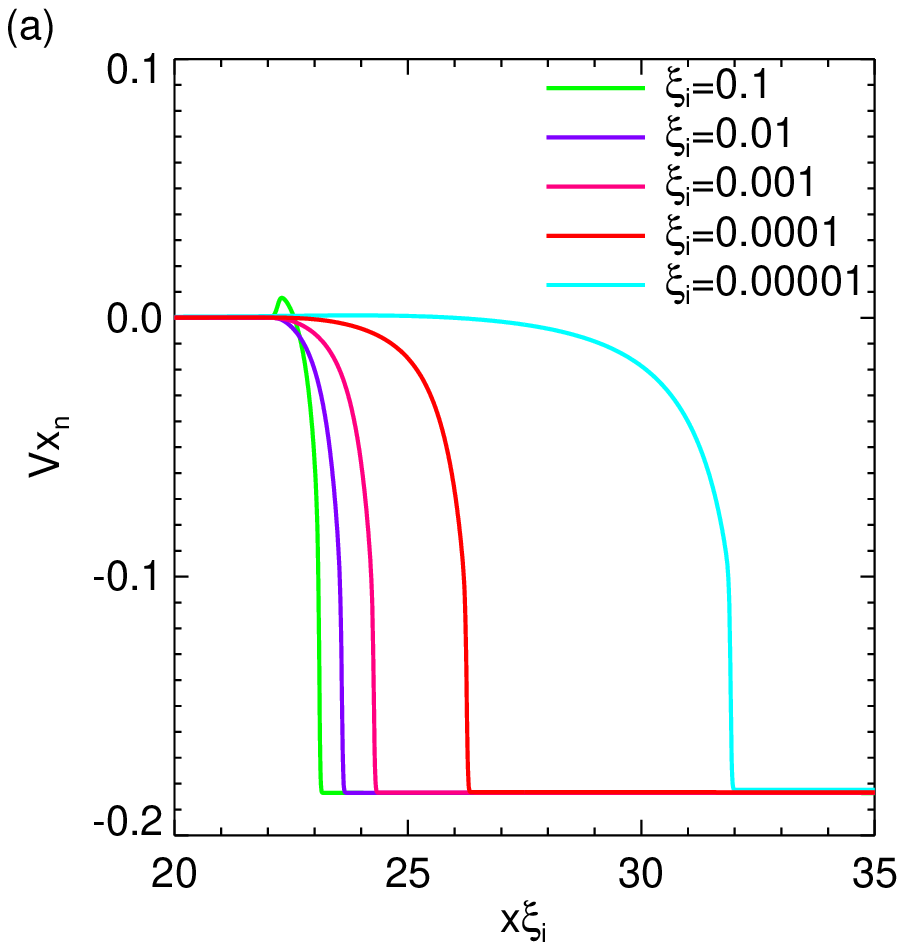}
\includegraphics[width=6.5cm]{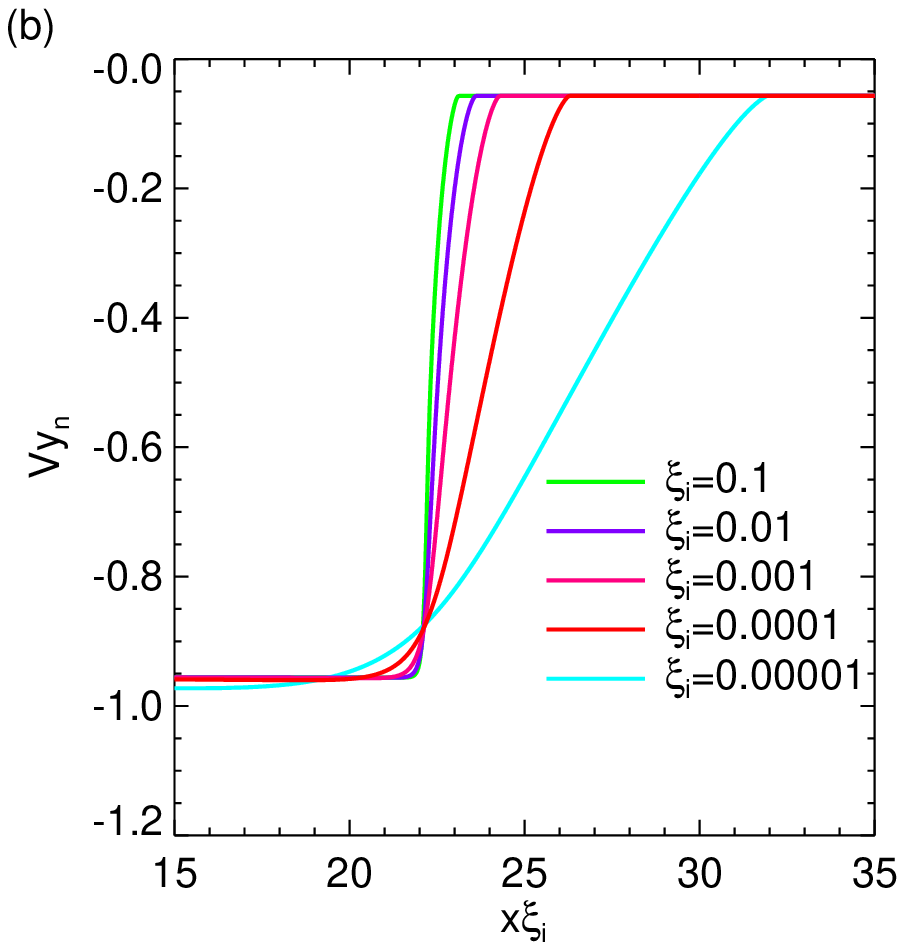}
\caption{Spatial distributions of $v_{\rm xn}$ (a) and $v_{\rm yn}$ (b) for $\beta=0.1$ and $\xi_{\rm i}=0.1$, $0.01$, $0.001$, $0.0001,$ and $0.00001$ denoted by the green, purple, pink, red, and turquoise lines, respectively.}
\label{shock_struc}
\end{figure*}

Figure \ref{shock_struc} gives the velocity distribution in the $x$ (panel a) and $y$ (panel b) directions for the case where $\beta=0.1$ and $\xi_{\rm i}$ varies between $0.00001$ and $0.1$.
The time that the snapshot is taken from each simulation is given by $t_{\rm snap}= 150\tau/\xi_{\rm i}$.
The axis of this figure is given as $x\xi_{\rm i}$.
This is because the collision frequency of the neutrals to the ions is proportional to the ionisation fraction and so it could be simply expected that the thickness of the shock region ($T_{\rm shock}$) would roughly scale as $T_{\rm shock}=V_{\rm shock}/\alpha_{\rm c}\rho_{\rm tot}\xi_{\rm i}$ \citep[e.g.][]{DRAINE1993}.
It is clear that this is not the case, but interestingly, for the $v_{\rm x}$ velocity the exit point of the shock position is a point of the system that evolves self-similarly (see Fig. \ref{QSS}); therefore, the time scaling used for the snapshot multiplied by $\xi_{\rm i}$ relates to the self-similar evolution of the system and that appears to be independent of ionisation fraction and hence aligns in this figure.
This is not the case for the $v_{\rm y}$ velocity as both the entrance and exit points for the fluid from the shock move further outwards as the ionisation fraction decreases.

It is also clear in Fig. \ref{shock_struc} that for the ionisation fraction of $\xi_{\rm i}=0.1$ there is an {overshoot in} the neutral velocity.
Though such features often appear in shock simulations as a result of numerical effects, this particular feature is very well resolved (approximately 100 grid points) and as such can be taken as a genuine feature of this particular shock.

\subsubsection{Structure of the drift velocity across the shock}

\begin{figure*}[ht]
\centering
\includegraphics[width=6.5cm]{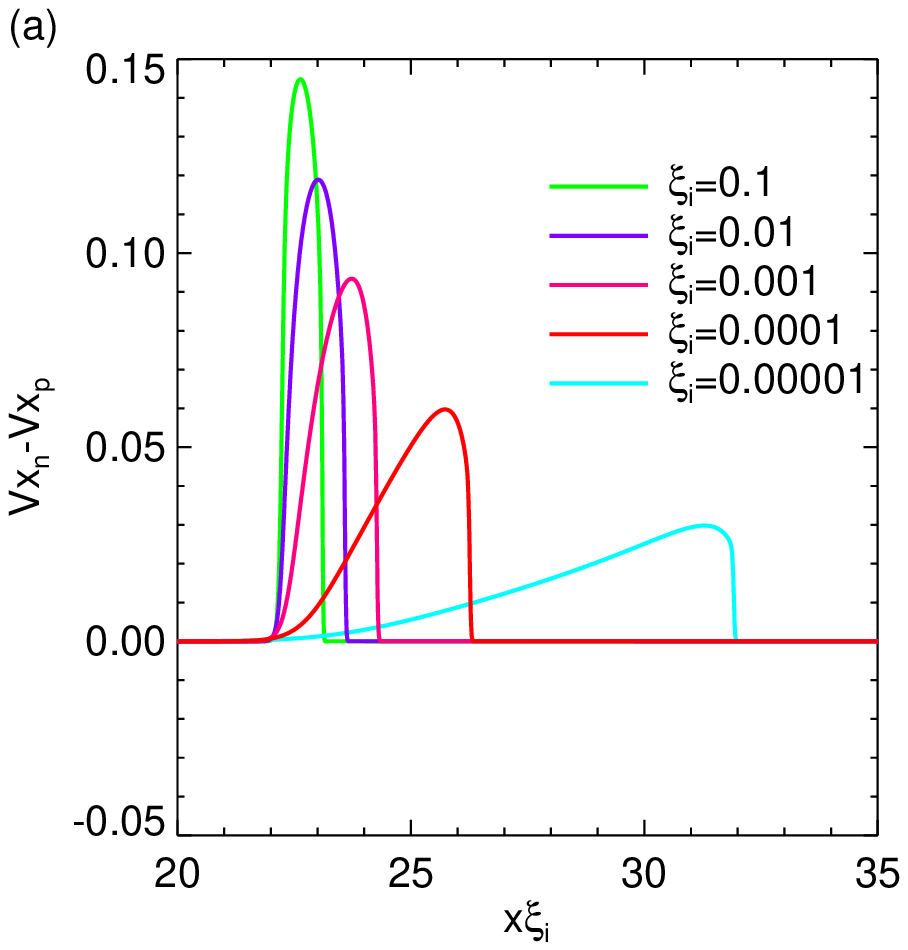}
\includegraphics[width=6.5cm]{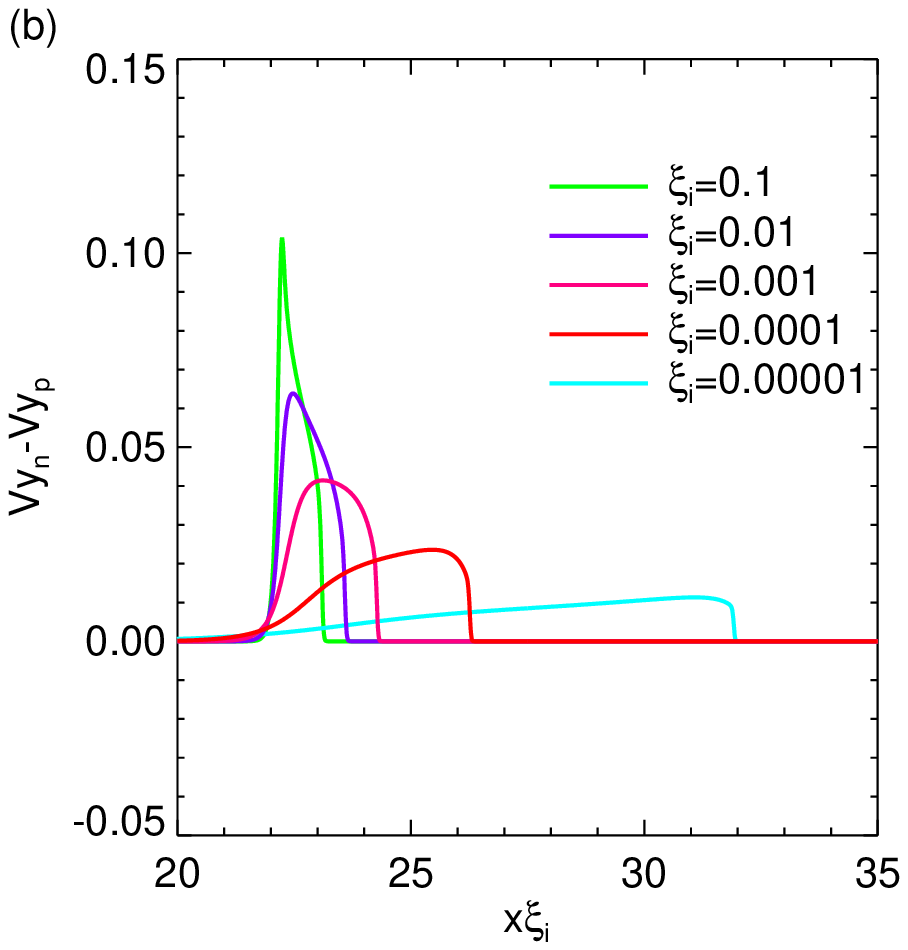}
\caption{Spatial distributions of $v_{\rm xn}-v_{\rm xp}$ (a) and $v_{\rm yn}-v_{\rm yp}$ (b) for $\xi_{\rm i}=0.1$, $0.01$, $0.001$, $0.0001,$ and $0.00001$. }
\label{shock_struc_drift}
\end{figure*}

Figure \ref{shock_struc_drift} gives the distribution of the drift velocity across the shock front for the range of ionisation fractions under study.
For large values of $\xi_{\rm i}$ the drift velocity profile for the $x$ velocities is approximately a Gaussian distribution, but as $\xi_{\rm i}$ gets smaller, the distribution changes and it develops a strong skew with the transition at the entry to the shock region becoming associated with a significantly sharper transition than the exit from the shock region.
The $y$-direction velocities with the drift velocity felt by a fluid packet steadily increasing as it moves through the shock before suddenly returning to $0$ for the case of $\xi_{\rm i}=0.1$.
However, the skew of this distribution also becomes progressively more and more positive as the ionisation fraction is decreased.

\subsubsection{Thickness of the shock region}

\begin{figure}[ht]
\centering
\includegraphics[width=6.5cm]{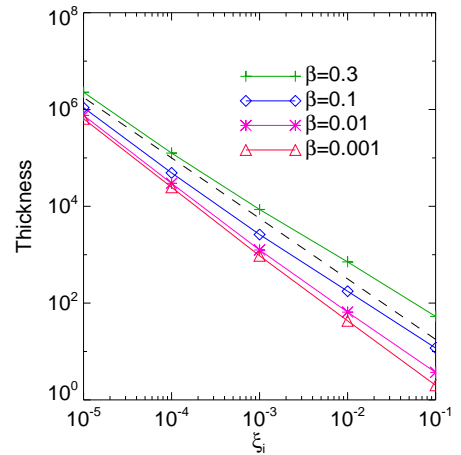}
\caption{Thickness of the shock region {as a function of} $\xi_{\rm i}$ for $\beta=0.3$ (green), $0.1$ (blue), $0.01$ (pink), and $0.001$ (red). The black dashed line shows a power law {as a function of} $\xi_{\rm i}$ with exponent $-1.2$}
\label{shock_thick_xi}
\end{figure}

The thickness of the shock region as a function of the ionisation fraction is given in Fig. \ref{shock_thick_xi}.
This thickness is calculated as the extent over which the drift velocity is creating momentum coupling in the $x$ direction across the shock front (it is clear from Fig. \ref{shock_struc_drift} that the values calculated from using the velocity drift in the $y$ direction would result in a different scaling). 
The scaling is given for $\beta=0.3$, $0.1$, $0.01,$ and $0.001$.

The general trend shown in this figure is that the thickness of the layer decreases monotonically with increase in the ionisation fraction.
However, as can be expected from the results in Fig. \ref{shock_struc} and \ref{shock_struc_drift}, the relation is not linear, but has a dependence that is approximately $\sim \xi_{\rm i}^{-1.2}$, as shown by the dashed line in Fig. \ref{shock_thick_xi}

\subsection{Plasma $\beta$ dependence}

\begin{figure*}[ht]
\centering
\includegraphics[width=6.5cm]{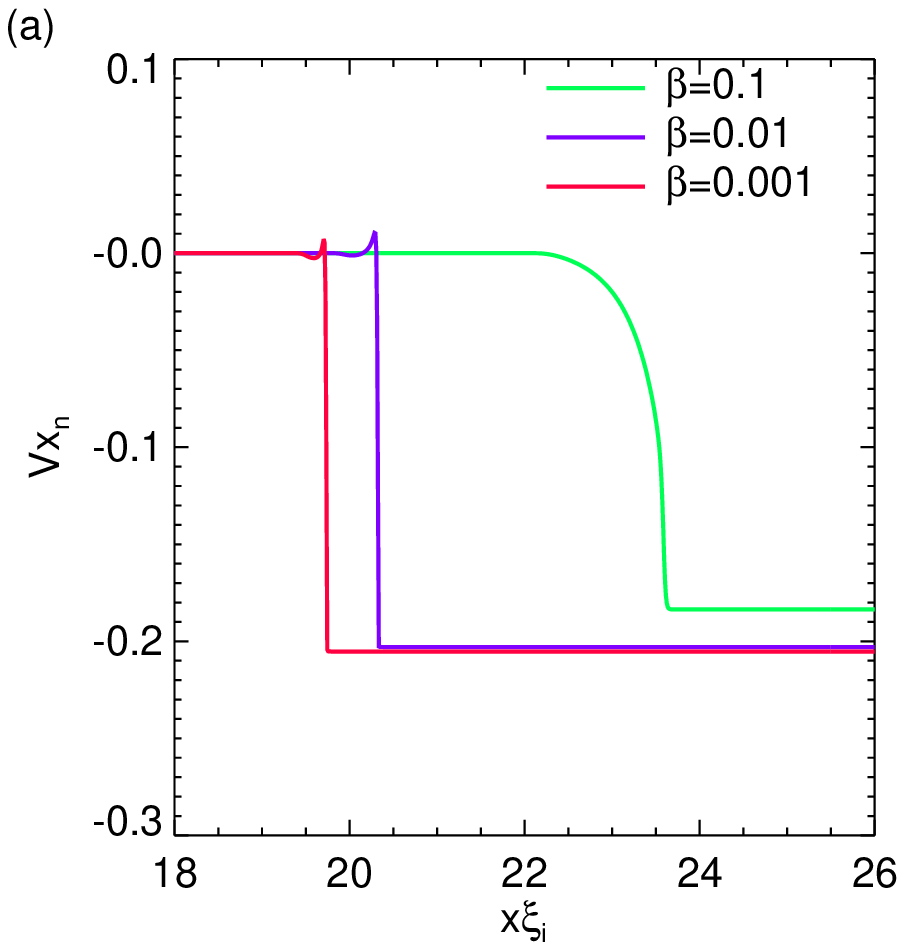}
\includegraphics[width=6.5cm]{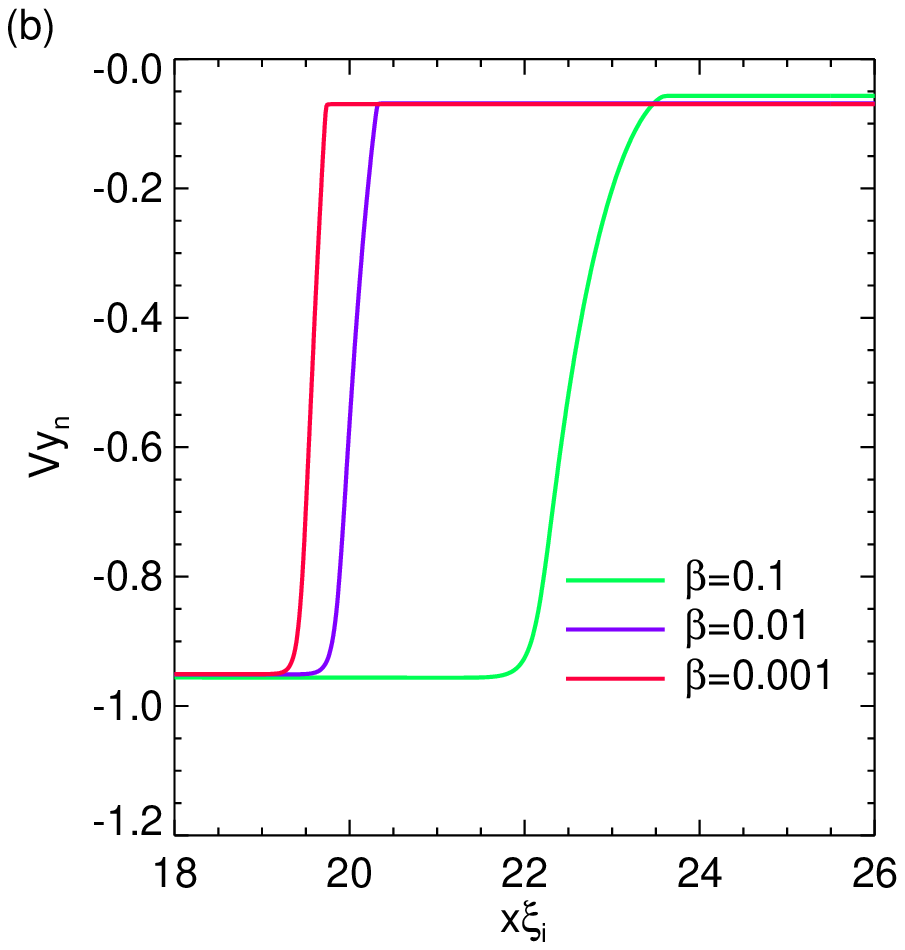}
\caption{Spatial distributions of $v_{\rm xn}$ (a) and $v_{\rm yn}$ (b) for $\xi=0.01$ and $\beta=0.1$, $0.01$ and $0.001$ denoted by the green red and blue lines, respectively.}
\label{shock_struc_beta}
\end{figure*}

The previous subsection dealt with the influence of the ionisation fraction; here we  now look at the dependence on the shock front of the plasma $\beta$.
Figure \ref{shock_struc_beta} is the same as Fig. \ref{shock_struc} but the different curves represent different plasma $\beta$ values instead of different $\xi_{\rm i}$ values.
We are looking at the case of $\xi_{\rm i}=0.01$.
The key point of this figure is that as plasma $\beta$ gets smaller, the system undergoes a transition from producing c-shocks to that of producing j-shocks where there is the jump in physical variables associated with a shock (in the $x$ velocity and also the pressure and density, but not the $y$ velocity).
The physics behind this are investigated in a following subsection.

It can be expected that the overshoot in the $x$ velocity (something that is resolved by approximately 100 grid points) is a real feature of the system.
In fact, we can estimate the downstream velocity of this shock by just performing a hydrodynamic analysis using the shock jump conditions on the shock, i.e. assuming that as this is a discontinuity {in which }the neutrals are not influenced by the ions and so the shock jump conditions should be just those of the hydrodynamic Rankine-Hugoniot relation.
For an ideal gas with adiabatic index $\gamma=5/3$, the ratio of the {\rm upstream} and {\rm downstream} velocities in the shock frame is $u_{\rm U}/u_{\rm D} \rightarrow 4$ as the shock gets progressively stronger.
For the case of $\beta=0.01$, we have a shock velocity of $\sim 0.15V_{\rm A}$ and an inflow velocity in the simulation frame of $\sim -0.2 V_A$, which would give an upper estimate for the post-shock velocity in the simulation frame of $\sim 0.06 V_{\rm A}$, i.e. the {velocity undergoes a positive overshoot above the $v_x=0$ level }that is expected for the final state after the fluids have recoupled.
The actual post-shock velocities shown in Fig. \ref{shock_struc_beta} are $\sim 0.02V_{\rm A}$ so they are within the predicted limit.
Once the shock has passed, the overshoot in the velocity decays exponentially, through collisional coupling over approximately one coupling length scale for neutrals to couple to the ions towards the $v_{\rm x}=0$ value.

It is also worth pointing out that there is an inflow dependence on $\beta$, which is driven by the rarefaction {wave. The inflow depends }on the physics of expansion of the system and the position of the shock front {depends} on the plasma $\beta$.
It is likely that part of the change in position of shock front (i.e. propagation speed of the shock front) is a result of the Doppler shifting of the shock by the inflow.
Compression of the shock also gives an explanation for what is happening.

\begin{figure*}[ht]
\centering
\includegraphics[width=13.5cm]{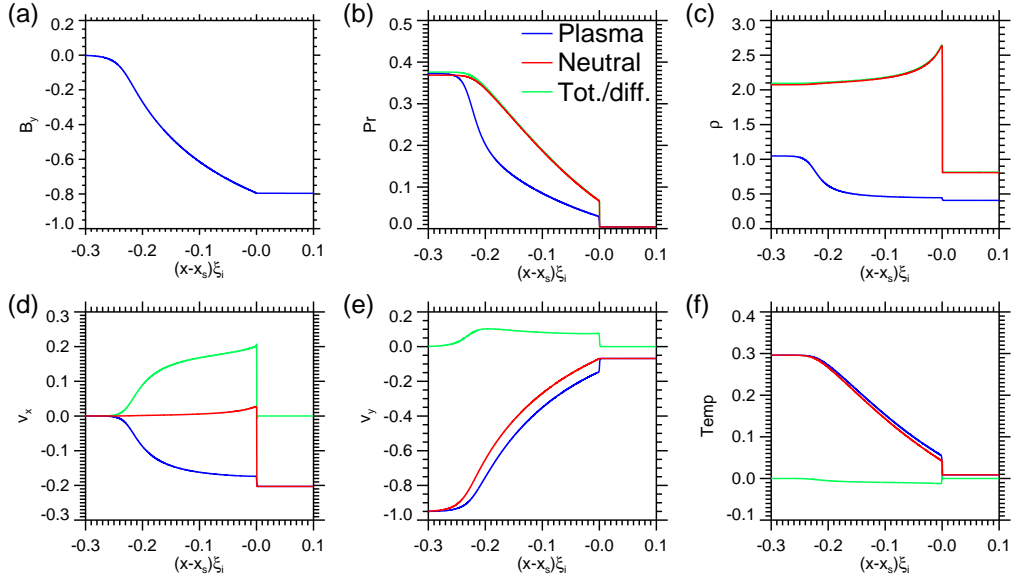}
\caption{Spatial distribution in the $x$ direction around the shock front and the coupling region of (a) $B_{\rm y}$, (b)  gas pressure, (c)  density, (d) $v_{\rm x}$, (e) $v_{\rm y}$ , and (f) the temperature for the neutral (red) and ionised (blue) fluids at time $t=15000 \tau$ where $\beta=0.01$ and $\xi_{\rm i}=0.01$. The green line indicates the total (pressure and density) or the difference ($x$ and $y$ velocities and temperature) for the two fluids. The plasma pressure and density are increased by a factor of 50 to make their distributions clearly visible. The $x$-axis has been shifted to set the origin at the shock front.}
\label{steady_state_shocked}
\end{figure*}

Figure \ref{steady_state_shocked} focusses on the distributions for both the neutral and ionised fluid around the shock region for the case where $\beta=0.01$ and $\xi_{\rm i}=0.01$.
A sharp transition is clear in $v_{\rm x}$, $P,$ and $\rho$ for both the neutral and ionised fluid.
This is a distinct difference from the j-shock solution as presented in Fig. 3 of \citet{DRAINE1993}, where the shock is only present in the neutral fluid because for this case when there is a j-shock in the neutral fluid, it is also present in the ionised fluid.
The pressure evolution through the shock front is rather interesting.
Though the neutral fluid shocks, where the kinetic energy transfers to thermal energy, most of the energy conversion goes from magnetic energy to thermal energy and kinetic energy (for the velocity in the $y$ direction).
Therefore, even after the shock, the pressure steadily increases.
The density overshoots the post-shock value as the neutral fluid shocks, but because of the diverging neutral velocity field the density decays to the post-shock value.

\subsubsection{Structure of the drift velocity across the shock}

\begin{figure*}[ht]
\centering
\includegraphics[width=6.5cm]{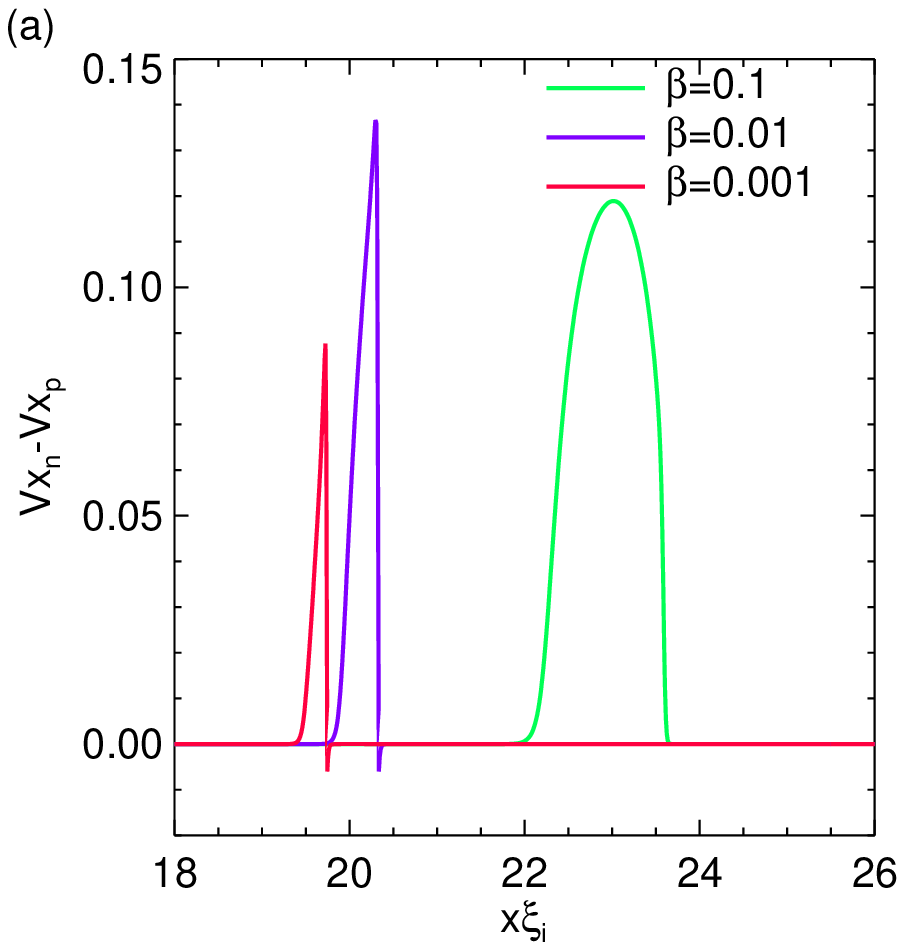}
\includegraphics[width=6.5cm]{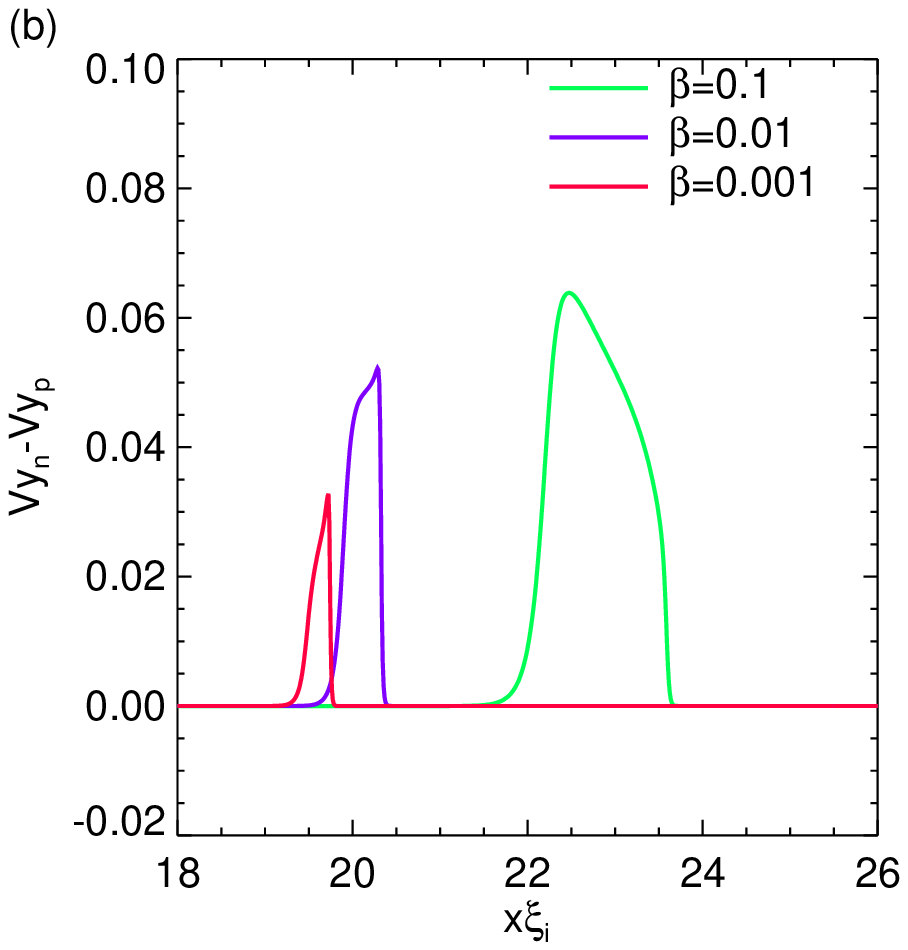}
\caption{Spatial distributions of $v_{\rm xn}-v_{\rm xp}$ (a) and $v_{\rm yn}-v_{\rm yp}$ (b) for $\beta=0.1$, $0.01$ and $0.001$ denoted by the green red and blue lines, respectively. }
\label{shock_struc_drift_beta}
\end{figure*}

Figure \ref{shock_struc_drift_beta} shows the drift velocity distributions for both the $x$ and $y$ velocities for different plasma $\beta$ values.
Once the shock has transitioned from a c-shock to a j-shock, the structure of the drift velocity is characterised by the sharp jump of the shock and a monotonic drop towards the $0$.

\subsubsection{Thickness of the shock region}

\begin{figure*}[ht]
\centering
\includegraphics[width=6.5cm]{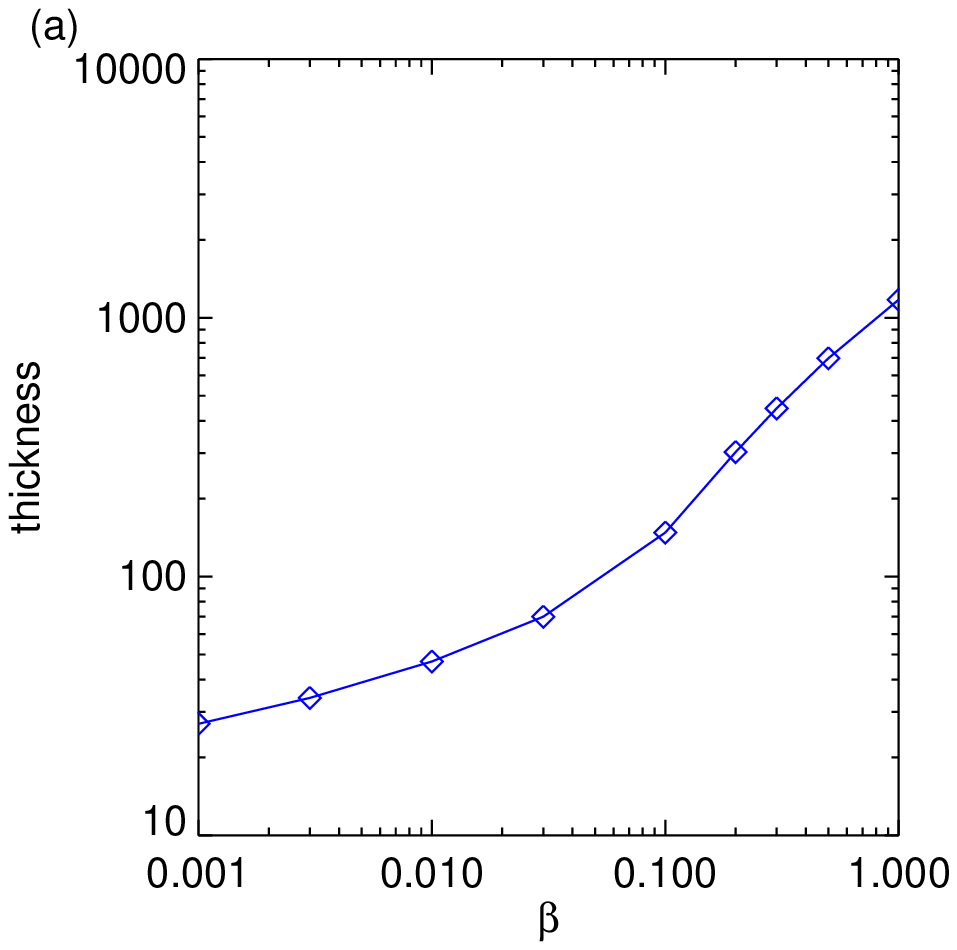}
\includegraphics[width=6.5cm]{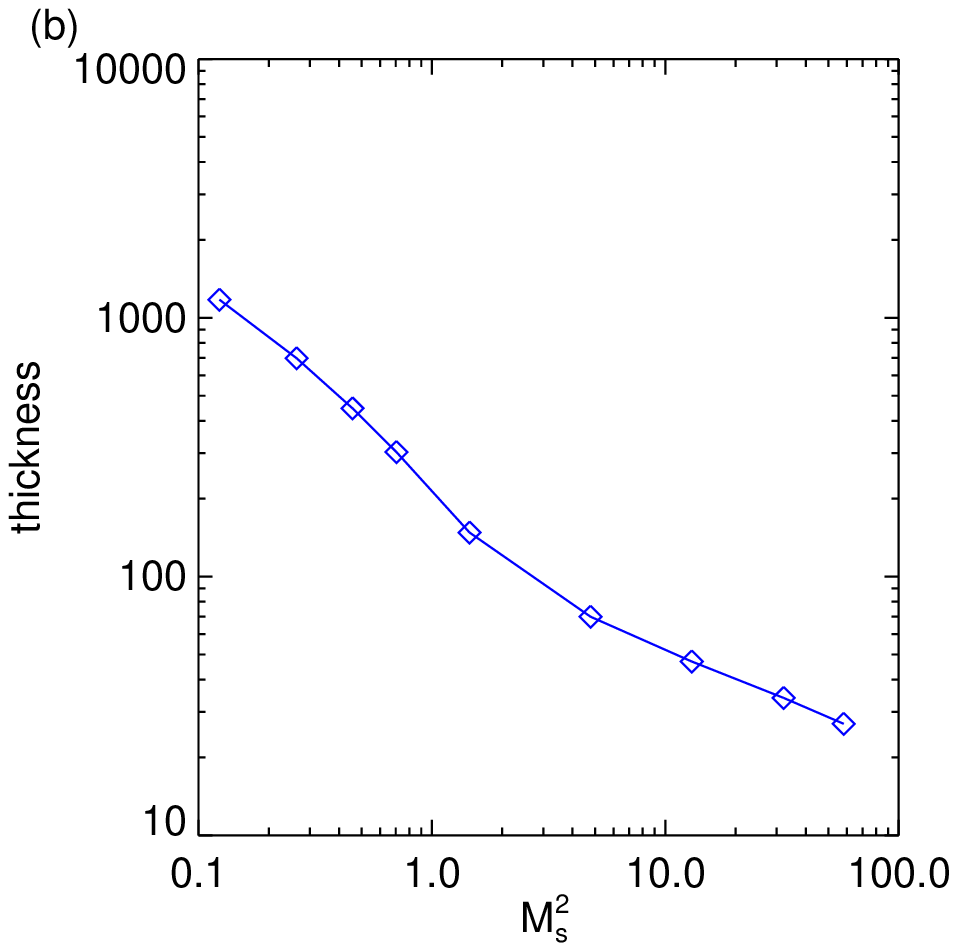}
\caption{Thickness (in units of $V_{\rm A}/\nu_{\rm T}$) of the shock for the simulations with $\xi_{\rm i}=0.01$ as a function of $\beta$ (left) and $M_{\rm n}^2$, the hydrodynamic Mach number of the neutral fluid in the inflow region, (right). The dependence of the thickness of the shock as a function of the plasma $\beta$ can be explained by the Mach number of the inflow region to which this relates to the compression at the shock front.}
\label{shock_thick_beta}
\end{figure*}

\begin{figure*}[ht]
\centering
\includegraphics[width=5.5cm]{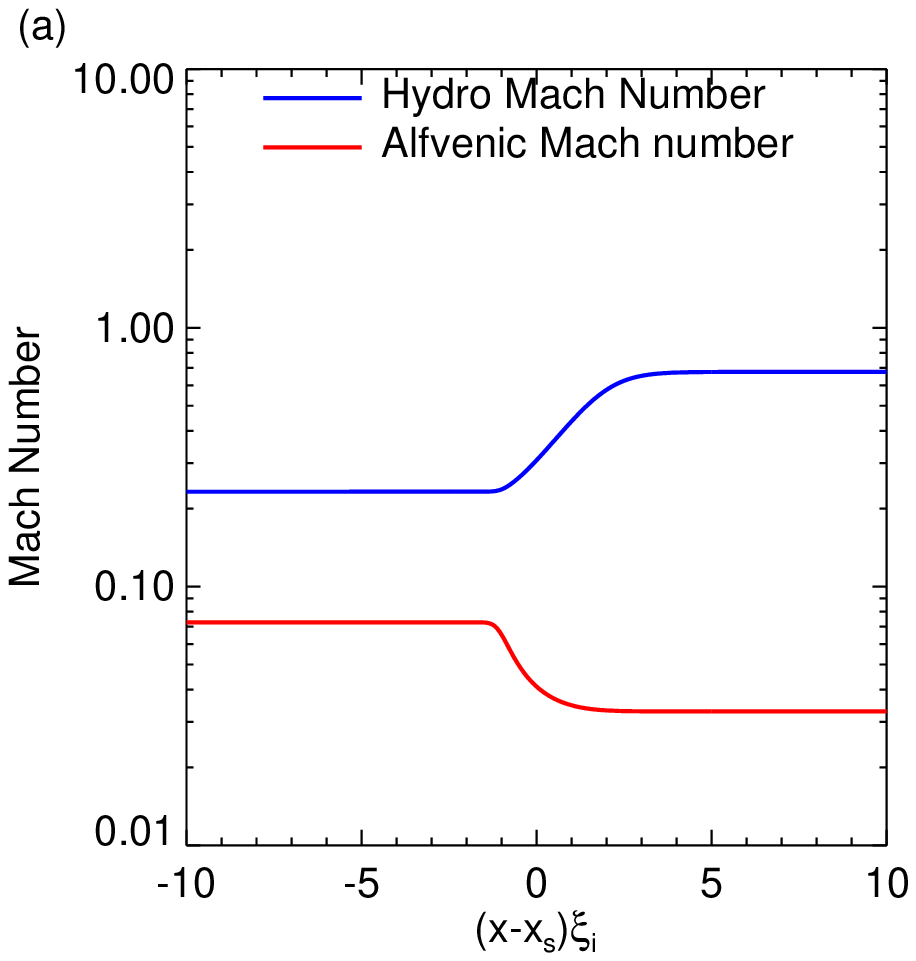}
\includegraphics[width=5.5cm]{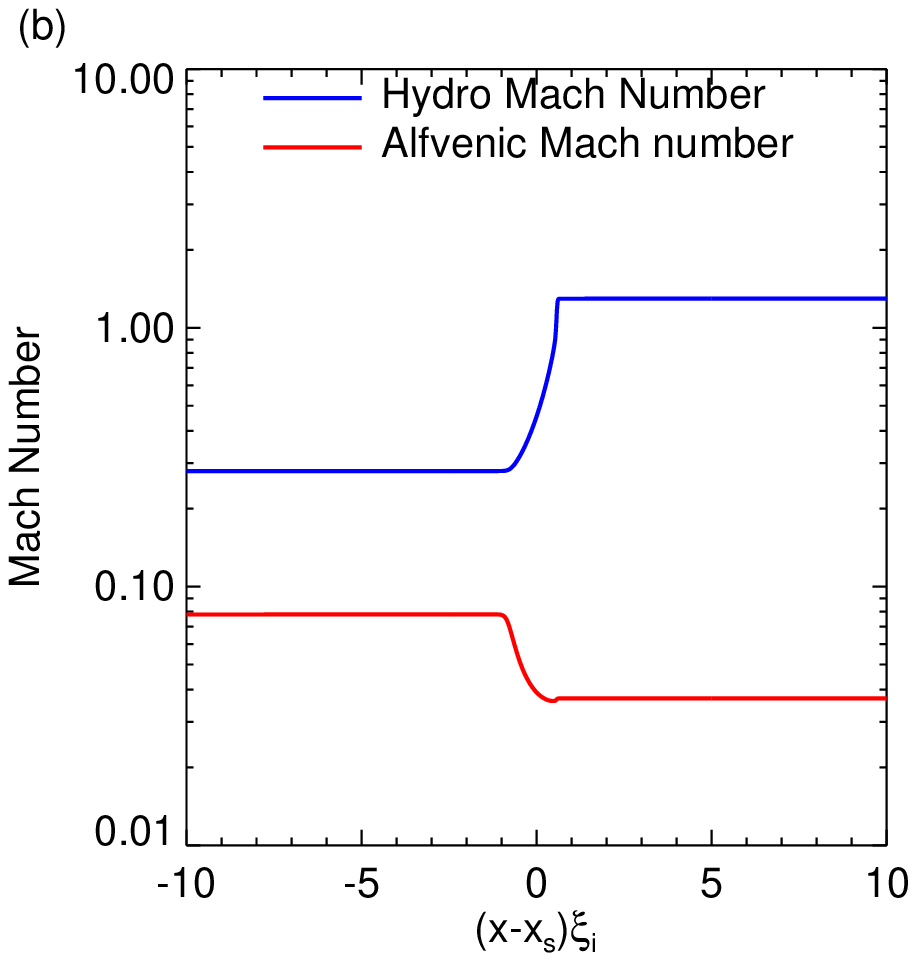}
\includegraphics[width=5.5cm]{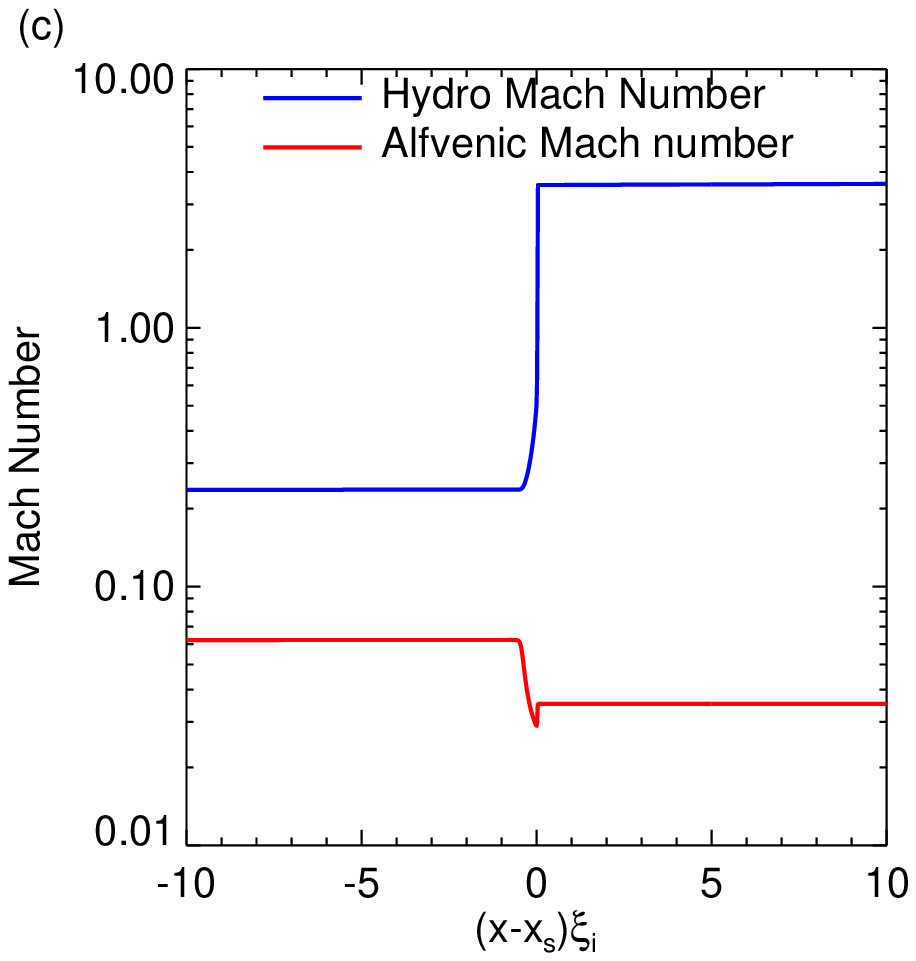}
\caption{Spatial distribution of $|v_{\rm xn}/Cs_{\rm n}|$ and $|v_{\rm xp}/V_{\rm A}|$ for a c-shock ($\beta=0.3$ and $\xi_{\rm i}=0.01$) in panel (a), a weak j-shock ($\beta=0.1$ and $\xi_{\rm i}=0.01$) in panel (b), and a j-shock ($\beta=0.01$ and $\xi_{\rm i}=0.01$) in panel (c) around the shock front (calculated in the shock frame). The $x$-axis is shifted by $x_{\rm s}$, where $x_{\rm s}$ is taken as the point in the shock where the drift velocity is at its peak, so that all distributions are aligned.}
\label{shock_struc_Mach}
\end{figure*}

Figure \ref{shock_thick_beta} gives the thickness of the shock region as a function of plasma $\beta$ (right) and the inflow Mach number of the neutrals in the shock frame $M_{\rm n}^2$ (left).
Though there is a monotonic increase with plasma $\beta,$ it does not become a clear power law.
To understand what happens when in the shock front {when $\beta$ changes,} it is worth looking at the upstream Mach number (calculated in the shock frame) of the neutral fluid.
Once the inflow speed exceeds the sound speed, a shock is formed (see discussion in next paragraph and Fig. \ref{shock_struc_Mach}); this limits the amount that it can be further compressed, so the nature of the coupling (where the temperature jump increases the collision frequency) then becomes important for determining how thick the layer becomes.

Figure \ref{shock_struc_Mach} shows the Mach number in the shock frame (calculated using $v_{\rm xn}$) of the neutral fluid and the Alfv\'{e}nic Mach number in the shock frame of the ionised fluid (calculated using $v_{\rm xp}$) for $\beta=0.3$, $0.1$ and $0.01$.
This choice of plasma $\beta$ gives a subsonic c-shock, a transonic weak j-shock, and a strong j-shock.
The $x$-axis is shifted by $x_{\rm s}$, where $x_{\rm s}$ is taken as the point in the shock where the drift velocity is at its peak, so that all distributions are aligned.
Here it is clear that for a j-shock to occur there has to be a supersonic inflow velocity, otherwise a c-shock forms.
For the case where the inflow velocity is only weakly supersonic, the shock is weak and the coupling across the shock front is very similar to that of the c-shock.
The Mach number drops to less than $0.3$ for all cases shown at post-shock .
At all times the ionised fluid is sub-Alfv\'{e}nic. 

\subsubsection{Fine structure of the j-shock}

\begin{figure*}[ht]
\centering
\includegraphics[width=13.5cm]{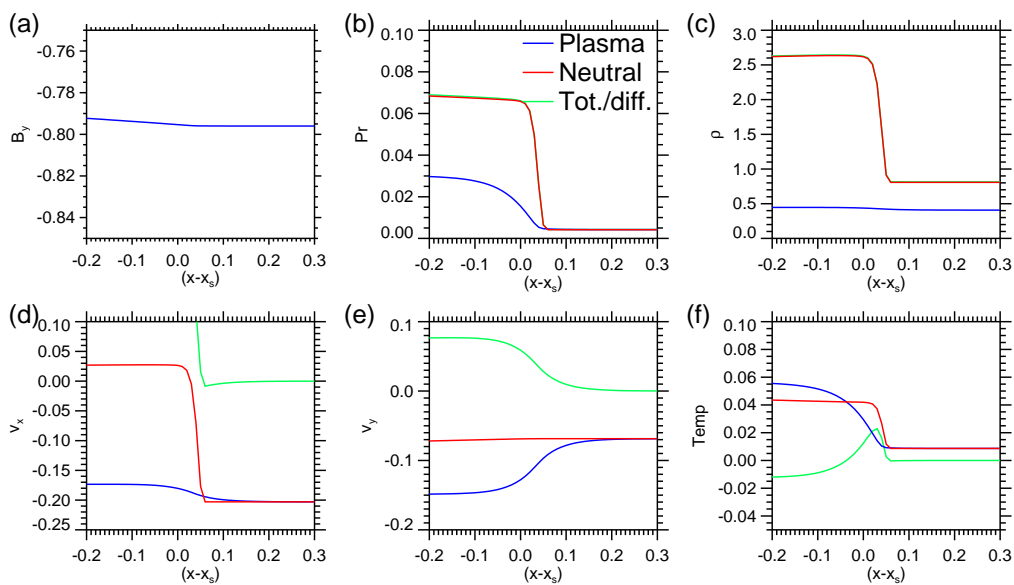}
\caption{Spatial distribution in the $x$ direction around the shock front of (a) $B_{\rm y}$, (b) the gas pressure, (c) the density, (d) $v_{\rm x}$, (e) $v_{\rm y}$, and (f) the temperature for the neutral (red) and ionised (blue) fluids at time $t=15000 \tau$ where $\beta=0.01$ and $\xi_{\rm i}=0.01$. The green line indicates the total (pressure and density) or the difference ($x$ and $y$ velocities and temperature) for the two fluids. The plasma pressure and density are increased by a factor of 50 to make their distributions clearly visible. }
\label{fine_struc}
\end{figure*}

In the shock frame, the ionised fluid is travelling towards the shock front, in this case at a velocity of $\sim 0.4V_{\rm A}$, so it can be expected that for a finite distance $D$ the ionised fluid could completely decouple from the neutral fluid around the shock front.
This distance can be estimated as $D=V_{\rm ion}/\alpha_{\rm c}\rho_{\rm n}$ and for the simulation of $\beta=0.01$ and $\xi_{\rm n}=0.99$ implies that $D$ is of order $0.1$.
Figure \ref{fine_struc} is a zoom in of the shock region of Fig. \ref{steady_state_shocked}.
Unlike Fig. \ref{steady_state_shocked}, for this calculation, $\Delta x =0.01,$ and {the units of the $x$-axis} of the figure have not been {rescaled} by $\xi_{\rm i}$ .

For the neutral fluid the shock transition happens over a few grid points determined by the numerical dissipation of the scheme, and the ionised fluid has a much {wider} transition.
The width of this transition is approximately the same length scale $D$ as estimated in the previous paragraph.

The finite width of the ionised fluid shock may have great importance for understanding how other diffusive and dissipative effects play a role in heating in the shock front.
If the thickness of the shock for the ionised fluid is determined by coupling processes then the Hall diffusion, magnetic diffusion, and viscosity would all be in regimes where {their} influence is small. As the ionised fluid maintains a finite thickness, it is likely that the neutral viscosity associated with the shock compression would become the other major dissipative process in the system.

\subsection{Frictional heating across the shock front}\label{heating}

\begin{figure*}[ht]
\centering
\includegraphics[width=6.5cm]{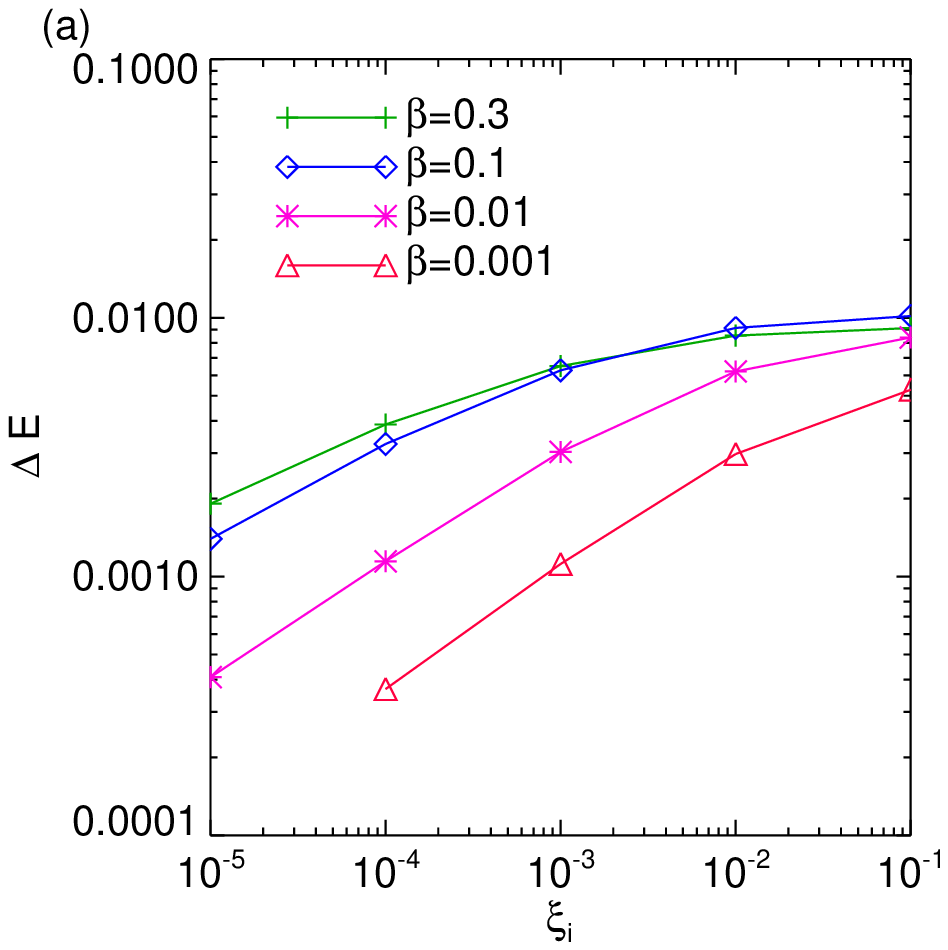}
\includegraphics[width=6.5cm]{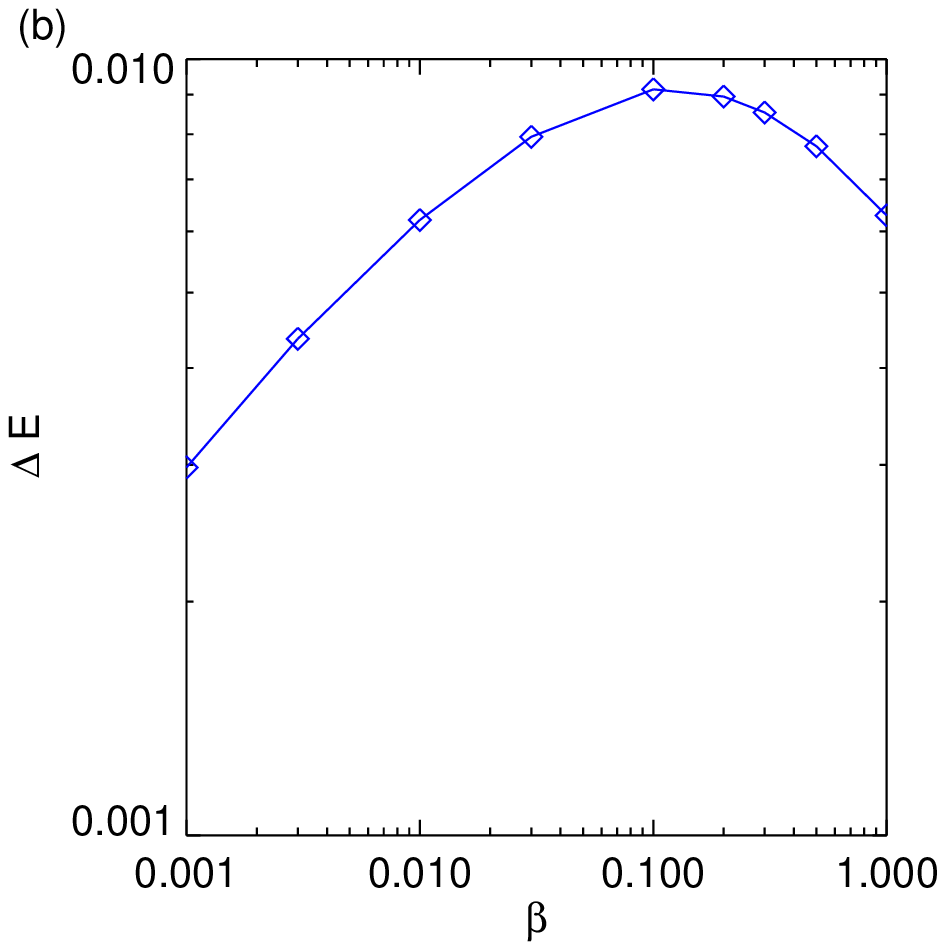}
\caption{Relation between the total energy that goes to heating the fluids through frictional heating as a package of fluid passes through the shock front for panel (a) $\xi_{\rm i}$ and panel (b) $\beta$ for $\xi_{\rm i}=0.01$.}
\label{shock_heat}
\end{figure*}

Figure \ref{shock_heat} indicates the total heating through frictional heating across the shock front as a function of $\xi_{\rm i}$ (panel a) and as a function of $\beta$ for $\xi_{\rm i}=0.01$ (panel b).
To give a context of the values shown, we note that the normalisation is the reference magnetic energy of $B_0^2/2=1/2$.
Therefore, the equivalent of a maximum of approximately 2 per cent of the reference magnetic energy is converted to heat in the shock front through the frictional heating process.

There are trends present in both the dependence on $\xi_{\rm i}$ and plasma $\beta$.
As the ionisation fraction decreases, the width of the shock front increases and, based on the arguments for the heating in the rarefaction wave presented in Sect. \ref{rarefaction}, we can expect that this leads to a decrease in the size of the nonlinear term resulting in a decrease in the overall heating.
For the plasma $\beta$, the heating peaks at $\beta=0.1$ (which is approximately where $M_{\rm n}^2=1$).
Because of the large temperature increase that results from the low $\beta$ shocks, the collisional frequency increases and this results in the reduction of the nonlinear terms across the shock front.

\section{Discussion}

In this study we investigated the 1D partially ionised MHD Riemann problem initiated by a magnetic slingshot in which the fully ionised ideal MHD version forms two waves: a fast-mode rarefaction wave and a slow-mode shock wave.
The partially ionised system {evolves} through four stages: the initiation, where most of the dynamics are in the ionised fluid and the coupling is mainly through the nonlinear energy terms; weak coupling, where the momentum coupling is beginning and the neutral dynamics are characterised by an explosive outflow; intermediate coupling, where the rarefaction wave and shock are forming as distinct entities in the system; and finally reaching a quasi-self-similar state.
In the quasi-self-similar stage, the shock forms a steady state but the rarefaction wave undergoes a diffusive evolution as it tends towards its final state.
A high-speed (approximately the Alfv\'{e}n velocity) jet forms behind the shock.
This is driven in the ionised fluid by the Lorentz force and in the neutral fluid by collisional coupling.

Two types of shock were found, a c-shock solution for low Mach number inflow and a j-shock solution when the inflow velocity becomes supersonic.
These are characterised by smooth transitions in the $x$ velocity, pressure, and density for both fluids in the c-shock and sharp transitions in both fluids for the j-shock.
Though in the model used the inflow is driven by a fast-mode rarefaction wave, which, as it is a compressional wave, gives a smaller inflow velocity when it can expand in three-dimensions (3D) than in 1D, so the dependence of the transition between the c- and j-shocks on the plasma $\beta$ may not be so applicable outside the model.
The dependence on the Mach number, however, is a universal rule because it relates exactly to the conditions of the shock jump, which is an inherently 1D problem.

One important point to discuss is the similarities and differences found in this study from those of shocks in the interstellar medium.
Though there are some large differences in which physical processes are under focus for each study, there is a great deal of comparison that can be made.
A key difference that should be highlighted is that because of the different geometries of the magnetic field under consideration, different shocks are formed.
Those under consideration in the ISM are fast-mode MHD shocks, where both the magnetic field strength and gas pressure increase downstream of the shock front, but we have been investigating a slow-mode MHD shock which through magnetic tension has a very large component of the downstream velocity parallel to the shock front.
The existence in both c- and j-shock solutions for both studies is one area of similarity, though the conditions for the appearance of the j-shock solution differ between the two studies.
For this study, j-shocks appeared based on the hydrodynamic Mach number of the {upstream flow: once this became supersonic, shocks formed}.
Charge exchange and ionisation/recombination were shown to strongly influence the shock structure in the ISM, and it would be no surprise if the shock studied here was the same.
For the influence of charge exchange it is likely that this process would act as an effective increase in the collisional coupling \citep[e.g.][]{TERR2015}.
It can be expected that, as the ionisation fraction is so important for determining the size of the coupling region around the shock, {ionisation associated with the shock} reduces the size of the shock region.

One application of this research is the investigation of shock-driven jets in the solar atmosphere.
Reconnection driving shock formation low in the solar atmosphere has been proposed as a key part of cool jet formation \citep[e.g.][]{SHIB1982, NAKA2012, TAKA2013}.
Our results suggest that owing to the changes in ionisation fraction and density with height in the solar atmosphere \citep{VAL1981} that any slow-mode shock that is propagating is constantly evolving as it travels through the atmosphere.
Our results also suggest that though limited in their spatial extent, observationally shock fronts, offer the greatest chance of observing ion-neutral drift in the solar {atmosphere as} shocks pass through the partially ionised regions of the atmosphere.

One interesting thought relates to the large gradients in $v_{\rm yn}$ with $x$ that result from the magnetic field driving the plasma jet.
Though {this strong} shear would only be experienced by an individual fluid packet for a finite amount of time as it passes through the shock front, there is the potential that it could have some interesting dynamic consequences. 
There is the possibility that it could drive the formation of velocity shear driven instabilities in the neutral fluid, which would be associated with turbulence in the neutral fluid in the reconnection jet and potentially thickening the velocity transition region associated with the shock.
It would be an interesting topic to investigate further.

\begin{acknowledgements}
A.H. is supported by his STFC Ernest Rutherford Fellowship grant number ST/L00397X/1.
S.T. acknowledges support by the Research Fellowship of the Japan Society for the Promotion of Science (JSPS).
The authors would like to thank Drs Hiroaki Isobe, Kazunari Shibata, Alexander Russell, \& David Chernoff for their insightful observations regarding this research.
This work used the DiRAC Data Centric system at Durham University, operated by the Institute for Computational Cosmology on behalf of the STFC DiRAC HPC Facility (www.dirac.ac.uk. This equipment was funded by a BIS National E-infrastructure capital grant ST/K00042X/1, STFC capital grant ST/K00087X/1, DiRAC Operations grant ST/K003267/1 and Durham University. DiRAC is part of the National E-Infrastructure.
\end{acknowledgements}

\begin{appendix}
\section{The (P\underline{I}P) code}\label{Appen1}

The (P\underline{I}P) code is a 3D numerical code designed to investigate the dynamics of partially ionised plasma across a range of spatial and temporal scales that can occur in astrophysical systems.
In this section we detail the full set of equations used in the code and a method to solve the collisional coupling terms to reduce the constraint they can have on the timestep of a simulation that is similar to the method presented in \citet{INOINU2008}.

\subsection{Equations}

In this code, two sets of equations are solved separately, which are then coupled using collisions, ionisation, and recombination. 
The equations used are those of an MHD fluid and a hydrodynamic (HD) fluid that are joined through collisional coupling.
The subscripts ${\rm p}$ and ${\rm n}$ are used to denote the variables for the MHD and HD equations, respectively.
The formulation of these terms can be found in \citet{LEAKE2012}, with \citet{Meier} and \citet{BRAG1965} as supplementary material.
For all the equations, the collisional, ionsation, and recombination terms are included on the RHS of the equation. 

The full equations solved for the evolution of the neutral hydrogen fluid are written as\begin{align}
\frac{\partial\rho_{\rm n}}{\partial t}+&\nabla\cdot(\rho_{\rm n}\mathbf{v}_{\rm n})=\gamma_{\rm rec}\rho_{\rm p}-\gamma_{\rm ion}\rho_{\rm n} \label{n_cont_eqn}\\
\frac{\partial}{\partial t}(\rho_{\rm n}\mathbf{v}_{\rm n})+&\nabla\cdot(\rho_{\rm n}\mathbf{v}_{\rm n}\mathbf{v}_{\rm n}+P_{\rm n}\mathbf{I})= \\
& \rho_{\rm n}\mathbf{g}-\alpha_{\rm c}\rho_{\rm n}\rho_{\rm p}(\mathbf{v}_{\rm n}-\mathbf{v}_{\rm p})+\gamma_{\rm rec}\rho_{\rm p}\mathbf{v}_{\rm p}-\gamma_{\rm ion}\rho_{\rm n}\mathbf{v}_{\rm n} \nonumber\\
\frac{\partial e_{\rm n}}{\partial t}+&\nabla\cdot[\mathbf{v}_{\rm n}(e_{\rm n}+P_{\rm n})+\kappa\nabla T_{\rm n}]=\\
&\rho_{\rm p}\mathbf{g}\cdot\mathbf{v}-\alpha_{\rm c}\rho_{\rm n}\rho_{\rm p}\left[\frac{1}{2}(\mathbf{v}_{\rm n}^2-\mathbf{v}_{\rm p}^2)+3R_{\rm g}(T_{\rm n}-T_{\rm p})\right]\nonumber \\&+\frac{1}{2}\gamma_{\rm rec}\rho_{\rm p}\mathbf{v}_{\rm p}^2-\frac{1}{2}\gamma_{\rm ion}\rho_{\rm n}\mathbf{v}_{\rm n}^2\nonumber.
\end{align}

The full equations solved for the evolution of the charge neutral ion-electron plasma fluid are written as
\begin{align}
\frac{\partial\rho_{\rm p}}{\partial t}+&\nabla\cdot(\rho_{\rm p}\mathbf{v}_{\rm p})=-\gamma_{\rm rec}\rho_{\rm p}+\gamma_{\rm ion}\rho_{\rm n} \\
\frac{\partial}{\partial t}(\rho_{\rm p}\mathbf{v}_{\rm p})+&\nabla\cdot\left(\rho_{\rm p}\mathbf{v}_{\rm p}\mathbf{v}_{\rm p}+P_{\rm p}\mathbf{I}-\frac{\mathbf{BB}}{4\pi}+\frac{\mathbf{B}^2}{8\pi}\right)= \\
&\rho_{\rm p}\mathbf{g}+\alpha_{\rm c}\rho_{\rm n}\rho_{\rm p}(\mathbf{v}_{\rm n}-\mathbf{v}_{\rm p})-\gamma_{\rm rec}\rho_{\rm p}\mathbf{v}_{\rm p}+ \gamma_{\rm ion}\rho_{\rm n}\mathbf{v}_{\rm n}\nonumber\\
\frac{\partial}{\partial t}\left( e_{\rm p}+\frac{B^2}{8\pi}\mathbf{I} \right)+& \nabla\cdot\left[\mathbf{v}_{\rm p}(e_{\rm p}+P_{\rm p})+\frac{c}{4\pi}\mathbf{E}\times\mathbf{B}+\kappa\nabla T_{\rm p}\right]=\\ 
&\rho_{\rm p}\mathbf{g}\cdot\mathbf{v}+\alpha_{\rm c}\rho_{\rm n}\rho_{\rm p}\left[\frac{1}{2}(\mathbf{v}_{\rm n}^2-\mathbf{v}_{\rm p}^2)+3R_{\rm g}(T_{\rm n}-T_{\rm p})\right]\nonumber\\&-\frac{1}{2}\gamma_{\rm rec}\rho_{\rm p}\mathbf{v}_{\rm p}^2+ \frac{1}{2}\gamma_{\rm ion}\rho_{\rm n}\mathbf{v}_{\rm n}^2 \nonumber \\
\frac{\partial \mathbf{B}}{\partial t}+&\nabla \times (\mathbf{v}_{\rm p}\times \mathbf{B}+\eta\mathbf{J}+\eta_{\rm A}\mathbf{B}^2\mathbf{J}_\perp)=0.\label{ind_eqn}
\end{align}
Equation \ref{ind_eqn} contains the Ambipolar diffusion term ($\eta_{\rm A}\mathbf{B}^2\mathbf{J}_\perp$).
However, this term can only be activated when single fluid simulations are performed.
The thermal conduction terms are solved using the method detailed in \citet{TAKA2015}.
The ionisation and recombination rates are given by \citep{Jef68}
\begin{align}
  \gamma_{\rm ion}(T_{\rm e}, n_{\rm e})=& 2.7 f(\frac{E_0}{k_{\rm B} T_{\rm e}})^{-2} T_{\rm e}^{-3/2} e^{-E_0/kT_{\rm e}} n_{\rm e} \\
  \gamma_{\rm rec}(T_{\rm e}, n_{\rm e}) =& 5.6 f \times 10^{-16} (\frac{E_0}{k_{\rm B} T_{\rm e}})^{-2} T^{-3} n_{\rm e}^2.
\end{align}
Two numerical schemes have been implemented: a fourth-order four-Runge-Kutta central difference scheme \citep{VOG2005} and an HLL scheme \citep{Har83} with the HLLC scheme for the HD equations \citep{TORO1994} and the HLLD scheme for the MHD equations \citep{MIYO2005}.

\subsection{Time integration of collisional source terms}

We have implemented two methods for calculating the update in the conserved variables. The first method is a simple numerical integration with time of the source terms (used in this paper because the dynamic timescales are less than the collisional timescales) and the second is a method that employs an exact solution for the integration of the equations that follow (this is appropriate when the collision timescale is much smaller than the other timescales of the system). 
A similar method can be found in \citet{INOINU2008}.

First, we assume collisional coupling time is much smaller than dynamic and ionisation and recombination term, then Equations \ref{n_cont_eqn} to \ref{ind_eqn} become
\begin{align}
  \frac{\partial \rho_\beta}{\partial t}=&0 (\beta={\rm n,p})\\
  \frac{\partial{\bf v}_{\rm p}}{\partial t} =&\alpha_{\rm c} \rho_{\rm n} ({\bf v}_{\rm n}-{\bf v}_{\rm p}) \\
  \frac{\partial{\bf v}_{\rm n}}{\partial t}=&-\alpha_{\rm c}  \rho_{\rm p}({\bf v}_{\rm n}-{\bf v}_{\rm p}) \\
  \frac{\partial\epsilon_{\rm n}}{\partial t}  =& -{\bf v}_{\rm n}\cdot \alpha_{\rm c}\rho_{\rm p} \rho_{\rm n}({\bf v}_{\rm n}-{\bf v}_{\rm p})\\
&+0.5\alpha_{\rm c} \rho_{\rm p} \rho_{\rm n}({\bf v}_{\rm n}-{\bf v}_{\rm p})^2-3\alpha_{\rm c} \rho_{\rm p}\rho_{\rm n}R_{\rm g}(T_{\rm n}-T_{\rm p})\nonumber\\
  \frac{\partial}{\partial t}\left(\epsilon_{\rm p}+\frac{{\bf B}^2}{8\pi}\right) =&{\bf v}_{\rm p}\cdot \alpha_{\rm c}\rho_{\rm p} \rho_{\rm n}({\bf v}_{\rm n}-{\bf v}_{\rm p})\\
&+0.5\alpha_{\rm c} \rho_{\rm p} \rho_{\rm n}({\bf v}_{\rm n}-{\bf v}_{\rm p})^2+3\alpha_{\rm c} \rho_{\rm p} \rho_{\rm n} R_{\rm g} (T_{\rm n}-T_{\rm p})\nonumber\\
  \frac{\partial {\bf B}}{\partial t} =& 0\\
  \epsilon_\beta \equiv& 0.5 \rho_\beta {\bf v}_\beta^2 +\frac{\rho_\beta R_{\rm g} T_\beta}{(\gamma-1)\mu_\beta} (\beta ={\rm n,p}),
\end{align}
where $R_{\rm g}$ and $\mu_{\beta}$ are the gas constant and mean molecular weight.
We can solve these equations analytically.
By using this analytic solution, we can analytically integrate the equations over temporal spacing $\Delta t$ and the physical variables at time $t_0+\Delta t$ are given by
\begin{align}
  (\rho_{\rm n} {\bf v}_{\rm n})[\Delta t] =& \rho_{\rm n} {\bf v}_{\rm n}-\frac{\rho_{\rm n} \rho_{\rm p}}{\rho_{\rm p}+\rho_{\rm n}} {\bf w}_{\rm D}[0](1- \exp{(-\nu_{\rm col} \Delta t)})\\
  (\rho_{\rm p} {\bf v}_{\rm p})[\Delta t] =& \rho_{\rm p} {\bf v}_{\rm p}+\frac{\rho_{\rm n} \rho_{\rm p}}{\rho_{\rm p}+\rho_{\rm n}} {\bf w}_{\rm D}[0](1- \exp{(-\nu_{\rm col} \Delta t)})\\
  {\bf w}_{\rm D}[t] =& {\bf v}_{\rm n}[t] -{\bf v}_{\rm p}[t] \\
  \epsilon_{\rm n}[\Delta t] =& \epsilon_{\rm n} -S_{\rm np}(\Delta t)\\
  \epsilon_{\rm p}[\Delta t] =& \epsilon_{\rm p} +S_{\rm np}(\Delta t) \\
  S_{\rm np}(\Delta t) \equiv& \frac{A}{\nu_{\rm col}}(1-\exp{(-\nu_{\rm col} \Delta t)}) \\
&+\frac{B}{2\nu_{\rm col}}(1- \exp{(-2\nu_{\rm col}\Delta t)}) \nonumber \\
&+\frac{C}{\nu_{\rm thm}}(1-\exp{(-\nu_{\rm thm} \Delta t)})\nonumber\\
  A \equiv& \frac{\alpha_{\rm c} \rho_{\rm n} \rho_{\rm p}}{2}(v_{\rm n}^2-v_{\rm p}^2)\\
  B \equiv& -3 \alpha_{\rm c} \rho_{\rm n} \rho_{\rm p} R_{\rm g} \Delta T_{\rm drift} \\
  C \equiv& 3 \alpha_{\rm c} \rho_{\rm n} \rho_{\rm p} R_{\rm g}(T_{\rm n}-T_{\rm p}+\Delta T_{\rm drift}) \\
  \Delta T_{\rm drift} \equiv& \frac{(\gamma-1)(\mu_{\rm n}/\rho_{\rm n}-\mu_{\rm p}/\rho_{\rm p}) \alpha_{\rm c} \rho_{\rm n} \rho_{\rm p}}{2(2\nu_{\rm col}-\nu_{\rm thm})R_{\rm g}}(v_{\rm n}-v_{\rm p})^2 \\
  \nu_{\rm col} \equiv& \alpha_{\rm c} (\rho_{\rm n}+\rho_{\rm p})\\
  \nu_{\rm thm} \equiv& 3(\gamma -1)\alpha_{\rm c}(\mu_{\rm p} \rho_{\rm n} + \mu_{\rm n} \rho_{\rm p}).
\end{align}
Assuming the partial derivative equations are in the form
\begin{equation}
  \frac{\partial {\bf U}}{\partial t} +\nabla \cdot {\bf F_{mhd}}({\bf U}) = {\bf S_{col}}({\bf U})
,\end{equation}
where ${\bf U}, {\bf F_{mhd}},$ and ${\bf S_{col}}$ are state vector, flux vector of MHD equations, and source vector of collisional coupling term.

By using the operator splitting method, the time integration is performed in the following two steps:
\begin{align}
  {\bf U}_{\rm mhd} =& {\bf U}_0+\Delta t {\bf R}_{\rm mhd}({\bf U}_0)\\
  {\bf U}_{1} =& {\bf U}_{\rm mhd}+\Delta t {\bf S}_{\rm col}({\bf U}_0)\\
  {\bf R}_{\rm mhd}({\bf U}) \equiv& \nabla \cdot {\bf F}_{\rm mhd}({\bf U}).
\end{align}

In the four-step Runge-Kutta time integration, to get updated ($t=t_0 +\Delta t$) state vector ${\bf U}_1$ from state vector ${\bf U}_0 (t=t_0)$, we use the scheme below as follows:
\begin{align}
  {\bf U}_{1/4} =& {\bf U}_0 +\frac{\Delta t}{4} {\bf R_{mhd}}({\bf U}_0)\\
  {\bf U}_{1/3} =& {\bf U}_0 +\frac{\Delta t}{3} {\bf R_{mhd}}({\bf U}_{1/4})\\
  {\bf U}_{1/2} =& {\bf U}_0 +\frac{\Delta t}{2} {\bf R_{mhd}}({\bf U}_{1/3})\\
  {\bf U}_{1} =& {\bf U}_0 +\Delta t {\bf R_{mhd}}({\bf U}_{1/2}).
\end{align}
If the collisional temporal scale is much smaller than the MHD temporal scale, the analytic integration of collisional term should interrupt Runge-Kutta time integration. Therefore, four-step Runge-Kutta time integration with analytic integration of collisional term becomes 
\begin{align}
  {\bf U}_0(1/4) =& {\bf U}_0(0)+\frac{\Delta t}{4}{\bf S_{col}}({\bf U}_0(0))\\  
  {\bf U}_{1/4} =& {\bf U}_0(1/4)+\frac{\Delta t}{4}{\bf R_{mhd}}({\bf U}_0(0))\\  
  {\bf U}_0(1/3) =& {\bf U}_0(1/4)+\frac{\Delta t}{12}{\bf S_{col}}({\bf U}_{1/4}))\\
  {\bf U}_{1/3} =& {\bf U}_0(1/3)+\frac{\Delta t}{3}{\bf R_{mhd}}({\bf U}_{1/4})\\
  {\bf U}_0(1/2) =& {\bf U}_0(1/3)+\frac{\Delta t}{6}{\bf S_{col}}({\bf U}_{1/3})\\
  {\bf U}_{1/2} =& {\bf U}_0(1/2)+\frac{\Delta t}{2}{\bf R_{mhd}}({\bf U}_{1/3})\\
  {\bf U}_{1} =& {\bf U}_0(1/2)+\Delta t{\bf R_{mhd}}({\bf U}_{1/2})+\frac{\Delta t}{2} {\bf S_{col}}({\bf U}_{1/2}).
\end{align}

\begin{figure}
\includegraphics[width=8cm]{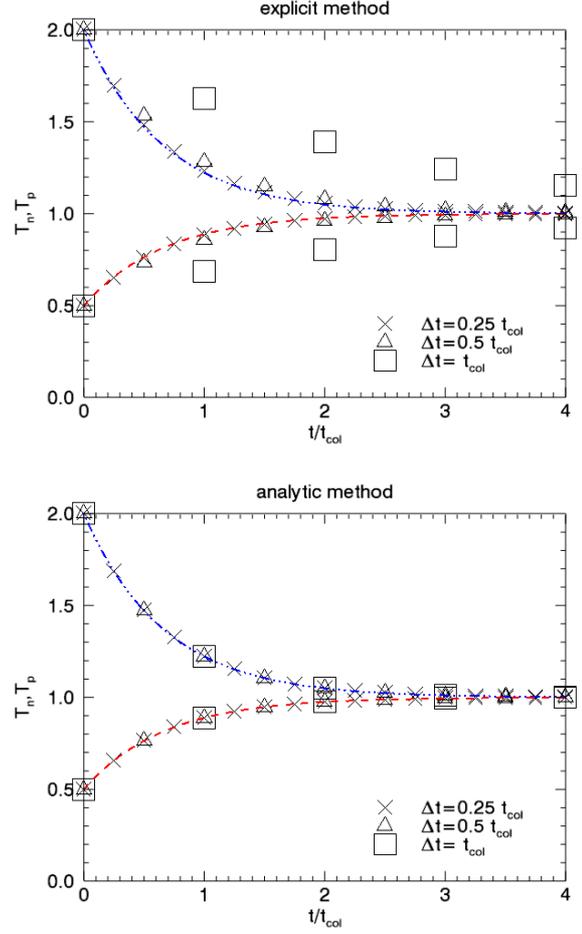}
\caption{Comparison between explicit method (upper panel) and analytic method (bottom panel). Three-dotted lines and dashed lines show analytic solution of neutral temperature and plasma temperature. Cross, triangle, and square symbols show numerical solutions with temporal spacing $\Delta t$ of 0.25, 0.5, and 1.0 of the collisional time $\tau_{\rm c}$.}
\label{analytic}
\end{figure}

Figure~\ref{analytic} shows the result of numerical test of our semi-analytic method. This is a 0-dimensional simulation to test the thermal coupling of the plasma and neutrals. At the start of the simulation, neutral temperature $T_{\rm n}=2$ and plasma temperature $T_{\rm p}=0.5$. This difference of temperature between neutrals and plasma decreases with time as a result of the thermal coupling terms in Equation (\ref{n_en}) and Equation (\ref{p_en}). {The upper panel of the figure shows the results of explicit methods using a temporal spacing that is comparable to the collisional timescale. The closer the timestep to the collsional timescale, the more the results deviate from the analytic solution.} On the other hand, the results of our semi-analytic method, shown in the lower panel of the figure, correspond to the analytic solution even if temporal spacing is comparable to the collisional timescale. Therefore, our semi-analytic method is useful for reducing the computational time by increasing the temporal spacing $\Delta t$ in a partially ionised plasma simulation. Our test suggest that time steps of up to $10\tau_{\rm c}$ can be accurately solved with this method. This method also has the great benefit of removing the numerical integration of the collisional terms, which are stiff and as such need a safety factor in the numerical integration of approximately 0.2 to maintain accuracy.

\end{appendix}
\end{document}